\documentclass[useAMS,usenatbib,times,a4]{mn2e}
\pdfoutput=1
\usepackage{aas_macros}
\usepackage{graphicx}
\usepackage{natbib} 
\usepackage[usenames]{color}

\def\apjs{ApJS }

\def\apjl{ApJL \rm}

\def\aap{AAP }

\def\Eq#1{{Eq.~(\ref{e:#1})}}% equation reference
\def\Fig#1{{Fig.~\ref{f:#1}}}% figure reference
              \def\bfx {{\bf x}} \def\bfn {{\bf n}}

\def\Fig#1{{Fig.~(\ref{f:#1})}}

\def\ee{\end{equation}} 
\def\ba{\begin{eqnarray}} 
\def\ea{\end{eqnarray}}

\def\bu{{\bf u}}   
      
    \def\bI{{\bf I}}
  \def\bu{{\bf u}}  
    \def\md{{\mathrm  d}}   \def\bF{{\bf F}}\def\bP{{\bf P}}\def\bD{{\bf D}}

\newcommand{\nicefrac}[2]{\leavevmode\kern.1em
            \raise.5ex\hbox{\the\scriptfont0                      #1}\kern-.1em
            /\kern-.15em\lower.25ex\hbox{\the\scriptfont0 #2}}

\title{A radiative transfer scheme for cosmological reionization based on a local Eddington tensor.}   
\author[Aubert \& Teyssier]
{Dominique Aubert$^{1,2,3}$\thanks{aubert@astro.u-strasbg.fr}, Romain Teyssier$^{2}$   \\   
$^{1}$Observatoire Astronomique de Strasbourg, 11  rue de l'Universite, 67000 Strasbourg, France\\  
$^{2}$Service  d'Astrophysique, CEA  Saclay,  Batiment 709, 91191 Gif  sur Yvette, France\\ 
$^{3}$Universite Louis Pasteur, Strasbourg, France}

\date{\today}

\begin{document}

\maketitle

\label{firstpage}

\begin{abstract}
A  radiative   transfer  scheme  is  presented,  based   on  a  moment
description of  the equation of radiative transfer  and the so--called
``M1 closure model''  for the Eddington tensor. This  model features a
strictly  hyperbolic  transport  step   for  radiation~:  it  has  been
implemented using standard Godunov--like techniques in a new code called
ATON.   Coupled   to  simple   models  of  ionization   chemistry  and
photo-heating, ATON is able to  reproduce the results of other schemes
on a  various set  of standard tests  such as  the expansion of  a HII
region,  the   shielding  of  the   radiation  by  dense   clumps  and
cosmological ionization by multiple sources.  Being simple yet robust,
such  a scheme  is intended  to be  naturally and  easily  included in
grid--based cosmological fluid solvers.
\end{abstract}


\section{Introduction}

During  the early  stages of  the Universe,  the process  of structure
formation  leads to the  formation of  the first  stars at  a redshift
$z\sim10-20$.   These  primordial,  metal-poor  stars  should  emit  a
substantial  amount of  radiation that  would  be able  to ionize  the
neutral    gas    that    fills    the    surrounding    space    (see
e.g. \citet{2001PhR...349..125B} for  an extensive review). Afterward,
it would become  transparent to radiation at high  energies, where the
transition  is commonly pictured  as ionized  bubbles that  expand and
percolate around sources. The investigation of the the distribution of
neutral gas at high redshift  is therefore a great source of knowledge
on the  first luminous  objects, on the  physical conditions  in which
they appear  and on  the cosmological context  that lead to  them. For
this purpose, several experiments were or are about to be set up : one
can mention LOFAR  or SKA which aim at detecting  the redshifted 21 cm
signal which would come up from the neutral hydrogen.

From a  theoretical point-of-view, the interest  of pursuing so-called
'full-physics' cosmological  simulations has been  recently emphasized
by  different   groups  %(e.g.   Horizon,  \citet{2006astro.ph..8289G},
%Ostriker et al.). 
These  numerical experiments can predict the history
of  star formation,  the large  scale gas  distribution  hosting small
scale disc--like objects, the  creation and ejection of metals, etc...
While  intensively tested at  low redshift,  only a  few observational
comparisons are available for $z>3$  and basically none for $z>5$.  In
this  context,  the  comparison  of  simulations'  prediction  to  the
observed 21 cm emission would open  a new range of epochs during which
the understanding of the formation of structures could be tested.

Common   features   of   hydrodynamical   simulations   are   gravity,
hydrodynamics and  star formation. They  are modeled self-consistently
and coupled to each other, up to a certain extent.  On the other hand,
the  radiation   is  often  considered  as   an  external  homogeneous
background and  its impact  on the physics  is not  exactly consistent
with the distribution  of sources inside the simulated  box.  A common
approach consists in taking  in account a diffuse radiation background
which  evolves  along  time  according  to  a  model  of  cosmological
radiation   sources   ( see e.g. \citet{1999ApJ...514..648M}).   The   geometrical
distribution  of  the  sources  and  the propagation  of  the  emitted
radiation  in the  neutral hydrogen  gas cannot  be eluded  if  one is
interested in the transition processes that occur at the reionization.
Note  that still  at lower  redshift, although  the universe  is fully
ionized, the  radiation field is  still highly inhomogeneous  down to
redshift 1 ( see e.g. \citet{1999ApJ...514..648M}).

A great  effort has been put  into developing tools  that simulate the
emission and  the propagation of  radiation in cosmological  boxes and
its impact  on the gas  physics through heating/cooling  processes and
ionization        (see        e.g.        \citet{2001NewA....6..437G},
\citet{2006MNRAS.372..679M}).  A perfect illustration of this interest
in  given  by  \citet{2006MNRAS.371.1057I}  where a  large  number  of
radiative transfer  codes were gathered in  order to be  tested on the
same  set of  numerical experiments.   Several methods  (ray shooting,
grid-based  codes, Monte-Carlo)  and several  types  of implementation
(post-processing, self-consistent  simulations) are compared  and this
project demonstrates  that different methods can  achieve very similar
results even though they greatly  differ in their conception. We would
like  to  stress  here  that   in  this  comparison  paper,  only  one
moment--based    code    was    used,    namely   the    OTVET    code
\citep{2001NewA....6..437G},  and  its  use  was  only  restricted  to
strictly periodic problems.

In this paper, we present  a new {\it moment--based} method to perform
calculations  of radiative transfer.  It relies  on a  rather standard
hyperbolic  grid--based solver  and  is simple  enough  to be  quickly
implemented.   This method  relies on  a momentum  description  of the
transfer equation  via the conservation of radiative  energy and flux.
\citet{2001NewA....6..437G} relied on the  same type of description of
the equation  of radiative transfer,  where the Eddington's  tensor is
constrained  by  the sources'  geometry,  assuming  an optically  thin
regime. We use here to  compute the Eddington tensor the so--called M1
closure  relation, which  provides  a variable  Eddington tensor  that
depends   only   on   the   local  radiation   flux   and   intensity.
Non-equilibrium ionization  is also taken  in account by  solving the
coupled set of  equations for out--of--equilibrium hydrogen chemistry.
The current  implementation performs only  post-processing of existing
simulation, even  though it  can in principle  be easily coupled  to a
grid-based hydrodynamical code.

The paper is organized as follows: the first section briefly introduces
the  model. The  numerical implementation  is then  presented  in more
details.  The third section is  devoted to the same tests performed in
\citet{2006MNRAS.371.1057I},  on rather academic  situations but also
on realistic  cosmological fields. The  range of applications  of this
new scheme is discussed in the last section.

\section{Radiative transfer as an hyperbolic system of conservation laws with source terms}

The current scheme is based on a momentum description of the radiative
transfer  equation. The  hierarchy of  equations is  truncated  at the
second order and the closure relation is provided by the M1 relation.
\subsection{Moments of the transfer equation}
The radiative transfer equation is given by: 
\begin{equation}
\frac{1}{c}\frac{\partial I_{\nu}}{\partial t} +\bfn\cdot{\bf\nabla} I_{\nu}=-\kappa_\nu I_\nu+\eta_\nu,
\label{eq:master}
\end{equation}
where  $I_{\nu}(\bfx,\bfn,t)$  is  the radiation  specific  intensity,
$\kappa_{\nu}(\bfx,\bfn,t)$    the    absorption    coefficient    and
$\eta_{\nu}(\bfx,\bfn,t)$  the  source function.  They  all depend  on
position, angle, frequency and time. The absorption coefficient, in the
context of ionizing radiation, is computed from
\begin{equation}
\kappa_\nu=n_{\rm H_0}\sigma_{\nu},
\end{equation}
where $\sigma_{\nu}$ is the photoionization cross section, and $n_{\rm
H_0}$ the neutral hydrogen density.
By taking  the first two  momenta of Eq. \ref{eq:master},  two coupled
equations can be obtained:
\begin{eqnarray}
\frac{\partial E_{\nu}}{ \partial t} +  \nabla \bF_{\nu} &=& -\kappa_{\nu}c E_{\nu} + S_{\nu},\label{e:mom1}\\
\frac{\partial \bF_{\nu}}{ \partial t} +  c^2\nabla \bP_{\nu} &=& -\kappa_{\nu} c \bF_{\nu}.\label{e:mom2} 
\end{eqnarray}
These  four equations  set the  conservation of  the  radiative energy
$E_{\nu}$,  the  zero-th order  momentum  of  the  intensity, and  the
conservation  of  the  radiative  flux $\bF_{\nu}$,  the  first  order
momentum.  The lower  dimensionality  of these  equations, plus  their
conservative   form,   make   them   more  suited   to   a   numerical
treatment. However, an expression  for the pressure tensor $\bP_{\nu}$
(i.e.  the second order momentum of the intensity) must be provided in
order  to  close  the   system  described  by  Eqs.  \ref{e:mom1}  and
\ref{e:mom2}. This issue is addressed in Sec. \ref{s:M1}.

From now  on, \Eq{mom1} and \Eq{mom2}  can be modified in  a form more
suitable to  the subsequent calculations.  First, the energy  and flux
densities  can  be replaced  by  \textit{number}  densities~: this  is
easily achieved by dividing \Eq{mom1} and \Eq{mom2} by a single photon
energy, $h\nu$. They become~:
\begin{eqnarray}
\frac{\partial N_{\nu}}{ \partial t} +  \nabla {\bF}_{\nu} &=& -\kappa_{\nu}c N_{\nu} + S_{\nu},\label{e:momn1}\\
\frac{\partial {\bF}_{\nu}}{ \partial t} +  c^2\nabla {\bP}_{\nu} &=& -\kappa_{\nu} c {\bF_{\nu}}\label{e:momn2} ,
\end{eqnarray}
where $N_{\nu}$ is the photon  number density. For sake of simplicity,
we  have used  the same  notation for  the photon  flux  (resp. photon
pressure  tensor) than  for  the energy  flux  (resp. energy  pressure
tensor), although they differ by the factor $h\nu$.
The source term is further divided into 2 contributions
\begin{eqnarray}
S_{\nu}=\dot N^*_\nu + \dot N^{rec}_\nu ~~~ {\rm where} ~~~
\dot N^{rec}_\nu = n_{\rm e} n_{\rm H_+} \dot \epsilon_\nu(T)
\end{eqnarray}
where the  first term is the  radiation coming from  stars or quasars
and the second term is the diffuse radiation due to recombination from
$\rm H_+$  .  Both radiation sources  are assumed to  be isotropic, so
that no source term appears in the flux equation. In the test section,
we will compare  our method to ray--tracing schemes  developed in the
context  of  cosmological  reionization \citep{2006MNRAS.371.1057I},  for  which
recombination   radiation  is   emitted  along   each  ray,   so  that
recombination radiation  is not isotropic  anymore. In order  to mimic
the  effect  of  ray--tracing,  we  optionally solve   a  modified
equation for the flux
\begin{eqnarray}
\frac{\partial {\bF}_{\nu}}{ \partial t} +  c^2\nabla {\bP}_{\nu} &=& -\kappa_{\nu} c {\bF_{\nu}}
+ \frac{\dot N^{rec}_\nu}{N_\nu}\bF_\nu\label{e:modflux} .
\label{e:rt1}
\end{eqnarray}
We call  this approximation the ``ray--tracing  scheme'', although the
underlying  method  still makes  use  of the  2  first  moments of  the
radiative transfer equation.

\subsection{Single group radiative transfer}

In the current implementation of ATON, we have restricted ourselves to
the  ionization  of a  single  specie,  namely  hydrogen, and  discard
completely  the fate  of helium  and other  elements. It  is relatively
straightforward  to extend  our scheme  to a  more  realistic chemical
composition,  using  for   example  the  multiple--frequency  approach
described in \citet{2001NewA....6..437G}. This  is beyond the scope of
this paper.  We  further simplify the problem by  considering only one
photon  group,  namely  all  photons  with  energy  greater  than  the
threshold  energy for  hydrogen.   We use  throughout  this paper  the
notations  introduced   by  \citet{1996ApJS..105...19K}.   The  number
density of  hydrogen nuclei is $n_{\rm  H}=\rho X /m_p$  (for which we
adopt $X=0.76$), while the  number density for neutral hydrogen (resp.
ionized hydrogen) is noted $n_{\rm H_0}$ (resp. $n_{\rm H_+}$).

We define the ionizing photons number density as
\begin{equation}
N_{\gamma \rm H_0}=\int^{\infty}_{\nu_{\rm H_0}}N_{\nu}\md\nu.
\end{equation}
Integrating \Eq{momn1} and \Eq{momn2} over photon frequency and dropping
the subscript $\rm H_0$ since we have only one photon group in this
paper, we get 
\begin{eqnarray}
\frac{\partial N_\gamma}{ \partial t} +  \nabla {\bf F}_\gamma &=& -n_{\rm H_0} c \sigma_N N_\gamma 
+ \dot N^*_\gamma + \dot N^{rec}_\gamma,\\
\frac{\partial {\bf F}_\gamma}{ \partial t} +  c^2\nabla {\bf P}_\gamma &=& 
- n_{\rm H_0} c \sigma_F{\bf F}_\gamma,
\end{eqnarray}
where we define two frequency--averaged cross--sections given by
\begin{equation}
\sigma_N N_\gamma = \int^{\infty}_{\nu_{\rm H_0}}\sigma_{\nu} N_{\nu}\md\nu~~~\rm{and}~~~
\sigma_F F_\gamma = \int^{\infty}_{\nu_{\rm H_0}}\sigma_{\nu} F_{\nu}\md\nu.
\label{crosssection}
\end{equation}
In order to simplify further the problem, we assume a simple reference
radiation intensity,  noted $J_0(\nu)$,  for which we  pre-compute the
average cross--section as follows
\begin{eqnarray}
\sigma_F \simeq \sigma_N \simeq \sigma_\gamma =   
\int^{\infty}_{\nu_{\rm H_0}}\sigma_{\nu} \frac{4\pi J_0(\nu)}{h\nu}\md\nu ~/
\int^{\infty}_{\nu_{\rm H_0}} \frac{4\pi J_0(\nu)}{h\nu}\md\nu,
\end{eqnarray}
where   the  hydrogen   photoionization   cross--section  $\sigma_{\rm
H_0}(\nu)$    is    taken    from    \citet{1997MNRAS.292...27H}.
Except in section \ref{s:fixed}, we will consider $10^5$ K black-body models where $\sigma_{\gamma}= 1.63 \times 10^{-18}$ cm$^2$.
%models:
%the power law case with $J_0(\nu) \propto \nu^{-\alpha}$ and the black
%body   case.   {\bf   ADD  SPECIFIC   VALUES  HERE}.    
Likewise,  the
recombination  radiation,  integrated  over  frequency in  our  single
energy group writes
\begin{eqnarray}
\dot N^{rec}_\gamma = \int^{\infty}_{\nu_{\rm H_0}} n_{\rm e} n_{\rm H_+} \dot \epsilon_{\rm H_+}(\nu,T)
= n_{\rm e} n_{\rm H_+} (\alpha_A-\alpha_B).
\end{eqnarray}
where $\alpha_A(T)$  (resp. $\alpha_B(T)$) is the case  A (resp.  case
B) recombination coefficient for $\rm  H_+$.  They are both taken from
from  \citet{1997MNRAS.292...27H}. The
set of equations we solve in this paper is finally~:
\begin{eqnarray}
\nonumber
\frac{\partial N_\gamma}{ \partial t} +  \nabla {\bf F}_\gamma = -n_{\rm H_0} c \sigma_\gamma N_\gamma 
&+& n_{\rm e} n_{\rm H_+} (\alpha_A-\alpha_B)\\
\label{e:ener}
&+& \dot N^*_\gamma ,\\
\frac{\partial {\bf F}_\gamma}{ \partial t} +  c^2\nabla {\bf P}_\gamma = - n_{\rm H_0} c \sigma_\gamma 
{\bf F}_\gamma.
\label{e:flux}
\end{eqnarray}
Let us  recall that this  set of equation  is obtained for  one single
group of photons, the ionizing ones. The same procedure can be
applied in  principle to  an arbitrary number  of groups, in  order to
achieve a better  spectral description of the problem.   The number of
systems to be  solved would therefore scale with  the number of groups
considered.

\subsection{Hydrogen thermochemistry}

In order  to close the last system  of equation, we need  to solve for
the time evolution of the  Hydrogen ionization fraction and of the gas
temperature. The chemical evolution of neutral hydrogen is governed by
a delicate balance between collisional ionization, photoionization and
collisional recombination.  These processes are part  of the following
evolution equation for $n_{\rm H_0}$
\begin{equation}
\frac{\rm D~ }{\rm D t}(n_{\rm H_0}) = \alpha_{A}n_{\rm e}n_{\rm H_+} 
-\beta n_{\rm e} n_{\rm H_0} - \Gamma_{\gamma \rm H_0} n_{\rm H_0},
\label{e:ion}
\end{equation}
together with charge conservation $n_{\rm e}=n_{\rm H_+}$ and Hydrogen
nuclei      conservation      $n_{\rm     H_+}+n_{\rm      H_0}=n_{\rm
H}$. $\Gamma_{\gamma  \rm H_0}$  is the Hydrogen  atom photoionization
rate,    given    by    (using    the   same    notations    as    for
Eq.~\ref{crosssection})
\begin{equation}
\Gamma_{\gamma \rm H_0} = c \sigma_\gamma N_\gamma.
\end{equation}
Radiative    cooling   and    photoionization    heating   are    also
self--consistently  taken  into account  by  solving  the gas  thermal
energy equation
\begin{equation}
\rho \frac{\rm D~ }{\rm D t}(\frac{e}{\rho})={\cal H}-{\cal L},
~~~{\rm with}~~~e=\frac{3}{2}n_{\rm tot}k_B T.
\label{e:temp}
\end{equation}
The cooling  rate, $\cal L$,  in erg/s/cc, is computed  using standard
collisional cooling processes due to case A and B recombination of Hydrogen, collisional ionization and excitation of Hydrogen and Bremsstrahlung. We use the cooling rates given by
\citet{1997MNRAS.292...27H},      \citet{2003MNRAS.345..379M}      and
references therein.  The photoionization  heating rate, $\cal H$, also
in erg/s/cc, is given by ${\cal H}=n_{\rm H_0}\dot \epsilon_{\rm H_0}$
where
\begin{equation}
\dot \epsilon_{\rm H_0}=c\int_{\nu_{\rm H_0}}
^\infty (h\nu-h\nu_{\rm H_0})\sigma_{\nu}N_{\nu}\md \nu=c\epsilon_\gamma\sigma_\gamma N_\gamma.
\end{equation}
Following the approach used in the last section, we approximate this
photoionization energy using the fiducial $10^5$ K black body
radiation  spectrum, so we  can precompute  the average  photon energy
$\epsilon_\gamma$ (equals to 29.65 eV in this case).   

%using   the aprop   fiducial   spectrum
%$J_0(\nu)$. {\bf ADD SPECIFIC VALUES HERE}.

\subsection{The M1 closure relation}
\label{s:M1}
As already  mentioned, we need to  specify the form  of the Eddington
tensor in order  to close the moment hierarchy  and solve the previous
set  of  equations.    \citet{2001NewA....6..437G}  suggested  to
compute  the Eddington tensor  assuming an  optically thin  medium and
summing up  the contribution of  all the background sources.   In this
way, the radiative intensity geometry was fully specified and the moment
equations could be solved. This techniques, refereed to as the ``Optically
Thin  Variable  Eddington Tensor''  method,  required  to  solve  four
different  Poisson-like equations. Since  all cosmology  codes already
have a  Poisson solver  to compute the  dark matter and  gas dynamics,
this  scheme  turned  out  to  be quite  efficient  and  accurate  for
cosmological  applications. In  this  paper, we  propose  to apply  a very simple closure relation to
cosmological reionization, called the
``M1 approximation'',  introduced more than  two decades ago  to solve
the radiative  transfer equations in  the optically thick  limit, while
retaining    some   accuracy   in    the   optically    thin   regime
\citep{1984JQSRT..31..149L}. 

It relies on the assumption that the radiation
angular distribution is axysymetric around the flux vector ${\bf F}$, 
so that the Eddington tensor, defined as $\bP=\bD N$, 
can be written in the general form:
\begin{equation}
\bD=\frac{1-\chi}{2}\bI+\frac{3\chi-1}{2}\bu\otimes\bu,
\label{e:etens}
\end{equation}
where $\bu$ is a unit vector  aligned with the flux direction. We need
also th define the reduced flux $\bf f$ by:
\begin{equation}
{\bf f} =\frac{\bF}{cN}=f \bu.
\label{e:rf}
\end{equation}
The  Eddington factor  $\chi$ is  a yet  unknown scalar  quantity that
depends only on $f$, and that should satisfy $1/3\le\chi\le1$.  We now
need a  model to  specify the functional  form for $\chi(f)$.   In his
review paper, \citet{1984JQSRT..31..149L} discussed a great variety of
closure relations.   The most simple {\it  and} physically meaningful
one is called the M1 model, for which we have:
\begin{equation}
\chi=\frac{3+4|{\bf f}|^2}{5+2\sqrt{4-3|{\bf f}|^2}}.
\end{equation}
This  closure relation corresponds  to the  angular distribution  of a
{\it        Lorentz       boosted}        isotropic       distribution
\citep{1984JQSRT..31..149L},  as for  the Cosmic  Microwave Background
dipole, for which the boost direction is aligned with the flux vector.
It has been shown by \citet{M1} that this closure relation is the only
one that minimize  the radiative entropy. 

This closure relation  is of course a very  crude approximation of the
true  radiation  distribution.   In  presence,  for  example,  of  two
distinct radiation source, the M1  model will replace the two sources
by  one  ``average'' source  in  between.   Nevertheless, this  colure
relation  has good  properties that  we now  discuss in  more details.
This  model satisfies  the physical  constraint  $1/3\le\chi\le1$. The
tensor $\bD$, which describes the radiation's local geometry, consists
in  two  separate  contributions.   The  first  one  is  an  isotropic
component, where the radiation affects all the directions in a similar
way.   In Eq.   \ref{e:etens}, the  second tensor  component  of $\bD$
exhibits principal  directions that are  aligned with the  local flux,
consistently  with a free-streaming  radiation. On  the one  hand, the
isotropic component disappears in  a pure transport regime with $f=1$,
i.e. $\chi=1$. On  the other hand, the diffusion  regime implies $f=0$
and $\chi=1/3$. Using this value  in Eq. \ref{e:etens} shows that only
the isotropic  component remains, as  expected. In a  general fashion,
all the intermediate regimes  represent local geometries where both an
isotropic radiation  and a free-streaming  radiation contribute. These
two limiting  cases are exactly described  by the M1  model, while the
general regimes are approximated by  a linear combination of these two
limiting cases.

The other interesting property is  that this model is purely local, so
that   no  expensive   Poisson  solvers   are  necessary.    Even  more
interestingly,   it  can  be   shown  that   the  left-hand   side  of
Equations~(\ref{e:ener}) and  (\ref{e:flux}) defines an  hyperbolic system
of  conservation laws,  with real  eigenvalues corresponding  to waves
traveling    at    (or    close    to)   the    speed    of    light.
\citep{M1,2007A&A...464..429G}.    We  can   therefore   use  standard
numerical  techniques designed  in the  general framework  of hyperbolic
conservation  laws  \citep{Toro97}  and  apply  them  to  cosmological
reionization.

\section {Numerical implementation}
\label{s:num}
We now  describe in details the  numerical scheme we  have designed to
solve the previous  set of equations. We use  the classical ``operator
splitting''  approach, decomposing  the equations  into  several steps
that are solved  in sequence. In the first  step, called the ``stellar
source step'', we add all ionizing photons coming from stellar sources
in the radiation  field.  In  the second step, called  the ``transport
step'', we  solve the hyperbolic system of  conservation laws described
in the  last section.  In the last  step, called  the ``thermochemical
step'', we solve the right-hand side of our radiation transport model,
together  with  the evolution  of  neutral  hydrogen  density and  gas
temperature.

For the first step, we perform in each cell of the computational grid,
indexed $i$, the following update:
\begin{eqnarray}
\left(N_\gamma \right)^{n+1}_i = \left(N_\gamma \right)^{n}_i 
+ \dot N^*_\gamma \Delta t
\end{eqnarray}
Here and in  the followings, index $n$ stands  for the radiation field
before the current step, and index $n+1$ for the radiation field after
the current step.  Since we have three intermediate steps, if we start
a given time $t$ with index $n$,  we reach the next time step at time
$t+\Delta t$ with index $n+3$.

\subsection{Transport Step}
The next operator we solve in our sequence is the hyperbolic system we have discussed
in the next section:
\begin{eqnarray}
\frac{\partial N_\gamma}{ \partial t} +  \nabla {\bf F}_\gamma = 0,\\
\frac{\partial {\bf F}_\gamma}{ \partial t} +  c^2\nabla {\bf P}_\gamma = 0,
\label{e:transport}
\end{eqnarray}
In  ATON, these equations can be solved either implicitly or explicitly, 
with the following integral form of the conservation laws (expressed here 
in 1D for sake of simplicity):
\begin{eqnarray}
\frac{\left(N_\gamma\right)^{n+1}_i-\left(N_\gamma\right)^{n}_i}{ \Delta t} + 
\frac{\left(F_\gamma\right)^m_{i+1/2}-\left(F_\gamma\right)^m_{i-1/2}}{\Delta x}&=& 0,\label{e:nexp}\\
\frac{\left(F_\gamma\right)^{n+1}_i-\left(F_\gamma\right)^{n}_i}{ \Delta t} + 
c^2\frac{\left(P_\gamma\right)^m_{i+1/2}-\left(P_\gamma\right)^m_{i-1/2}}{\Delta x}&=& 0.
\label{e:discrete}
\end{eqnarray}
The flux  function ${\cal F} = (F_\gamma,P_\gamma)^T$  is evaluated at
the intercell faces, indexed $i+1/2$,  and at time $m$, with $m=n$ for
an  explicit  scheme,  whose  stability  condition  imposes  a  strong
constraint on the  time step, or with $m=n+1$  for an implicit scheme,
unconditionnaly  stable.   Following  \citet{2007A&A...464..429G},  we
compute  the  intercell  flux  using  standard  methods  designed  for
computational  fluid dynamics  such  as the  Godunov  method or,  more
generally, such  as the  class of upwind  schemes.  If we  note ${\cal
U}=(N_\gamma, F_\gamma)^T$ the vector of state variable, the intercell
flux  depends  on  the left  and  right  states  with respect  to  the
interface:
\begin{eqnarray}
{\cal F}^m_{i+1/2} = {\cal F}({\cal U}^m_i,{\cal U}^m_{i+1}) 
\end{eqnarray}
\citet{2007A&A...464..429G}  have tested  various flux  functions with
respect to  the M1 model, and  came up with  two possibilities, namely
the Harten--Lax--van Leer (HLL) flux function, for which we have:
\begin{eqnarray}
\left({\cal F}_{HLL}\right)^m_{i+1/2} = 
\frac{ \lambda^+{\cal F}^m_{i}-\lambda^-{\cal F}^m_{i+1}+
\lambda^+\lambda^-({\cal U}^m_{i+1}-{\cal U}^m_{i})}
{\lambda^+-\lambda^-},
\end{eqnarray}
where $\lambda^+  = \max (0, \lambda_{max})$ and  $\lambda^- = \min(0,
\lambda_{min})$  are  the  maximum  and  minimum  eigenvalues  of  the
Jacobian matrix  of the  system evaluated at  the $i$-th  and $i+1$-th
cells   (for  more  details,   the  reader   is  encouraged   to  read
\citet{2007A&A...464..429G}) and  the Global Lax  Friedrich (GLF) flux
function for which the maximum wave  speed is taken equal to the speed
of light:
\begin{eqnarray}
\left({\cal F}_{GLF}\right)^m_{i+1/2} = 
\frac{ {\cal F}^m_{i}+{\cal F}^m_{i+1} }{2} -
\frac{c}{2} ({\cal U}^m_{i+1}-{\cal U}^m_{i}).
\end{eqnarray}
As  we  will demonstrate  in  the test  section,  the  GLF flux,  more
diffusive by nature, turns out to give results very similar to the HLL
flux, while being much simpler to implement, since it does not require
to determine the eigenvalues of a rather complex hyperbolic system.

\subsection{Implicit versus explicit ?}

The time step of our scheme,  $\Delta t$, is controlled by the Courant
condition, which writes $3 c\Delta  t/\Delta x <1$, if the integration
is performed  explicitly ($m=n$).  In cosmological  problems, this can
result  into very  small  time steps,  although  the overall  solution
evolves quite  slowly.  The  standard solution is  to use  larger time
steps, related for example to  the much lower ionization front typical
propagation speed, but we must rely on implicit integration, for which
$m=n+1$. In  the latter case, we  fix the Eddington  factor $\chi$ and
the various eigenvalues $\lambda$  to their initial value (index $n$),
so that  the implicit form of  our solver becomes a  linear system, as
can  be   seen  from  Equations~(\ref{e:etens}),   (\ref{e:nexp})  and
(\ref{e:discrete}).  This linear system  is then solved using standard
sparse  solvers, among which  the Gauss--Seidel  solver gave  the best
results.

Another alternative to overcome the limitation of the explicit scheme,
while avoiding the complexity of  the implicit scheme, was proposed by
\citet{2001NewA....6..437G}:     the     ``Reduced    Speed--of--Light
Approximation''. The  idea is  to replace the  actual speed--of--light
$c$  by an  effective speed--of--light  $\tilde c  \ll c$  in  all the
previous  equations  (including   the  photoionization  rates),  where
$10^{-3}<\tilde  c/c<1$  is  typically  considered.  By  reducing  the
speed-of-light, the  Courant condition becomes looser  and larger time
steps can  be considered. Even though it  introduces an approximation,
no consequences is found as long as phenomena where the actual speed
of propagation  remains smaller than  $\tilde c$ are  considered, like
the  expansion of  an HII  region. As  shown in  the  following, using
$\tilde c$  instead of $c$ has  no consequence on the  accuracy of the
results.  Furthermore,  there are also  a significant number  of cases
where  satisfying   the  strict  Courant  condition   can  be  handled
numerically, still avoiding the rather costly implicit solver. In this
paper,  we have  never used  the  implicit scheme,  although in  other
physical conditions, using the implicit solver can be unavoidable.

\subsection{Thermochemical step}

In  our operator  splitting  approach,  the last  step  solves for  the
chemical evolution of Hydrogen  and for the coupling between radiation
and matter.   Physical quantities such  as the gas temperature  or the
ionization fraction must  be updated. The equations we  solve here are
the followings:
\begin{eqnarray}
\frac{\partial N_\gamma}{ \partial t} &=& -n_{\rm H_0} c \sigma_\gamma N_\gamma 
+ n_{\rm e} n_{\rm H_+} (\alpha_A-\alpha_B)\\
\frac{\partial {\bf F}_\gamma}{ \partial t} &=& - n_{\rm H_0} c \sigma_\gamma {\bf F}_\gamma\\
\frac{\partial n_{\rm H_0}}{\partial t} &=& - n_{\rm H_0}c \sigma_\gamma N_\gamma 
+ n_{\rm e}n_{\rm H_+}\alpha_{A} - n_{\rm e} n_{\rm H_0}\beta,\\
\frac{\partial e }{\partial t} &=& n_{\rm H_0} c \sigma_\gamma \epsilon_\gamma N_\gamma 
-n_{\rm H_0}n_e\Lambda_{e{\rm H_0}}-n_{\rm H_+}n_e\Lambda_{e{\rm H_+}}.
\end{eqnarray}
In   these  equations,   the   coefficients  $\alpha_A$,   $\alpha_B$,
$\Lambda_{e{\rm   H_0}}$  and   $\Lambda_{e{\rm   H_+}}$  all   depend
non--linearly on  the gas  temperature. The various  chemical species'
densities can  be expressed as  a function of the  ionization fraction
$x$  as  $n_e=n_{\rm  H_+}=xn_{\rm  H}$ and  $n_{\rm  H_0}=(1-x)n_{\rm
H}$. The  gas internal energy is expressed  as $e=3/2(1+x)n_{\rm H}k_B
T$.  Because of  the very  small  time scales involved,  we solve  this
system  using a  fully implicit  scheme.  The  first two  equation are
linear  with respect  to  their  main unknown,  so  that the  implicit
discretization can be worked out analytically:
\begin{eqnarray}
\left(N_\gamma\right)^{n+1}&=&\frac{\left(N_\gamma\right)^{n}+\Delta t n_{\rm e}^{n+1} n_{\rm H_+}^{n+1} (\alpha^{n+1}_A-\alpha^{n+1}_B)}
{1+\Delta t n_{\rm H_0}^{n+1} c \sigma_\gamma},\\
\left({\bf F}_\gamma\right)^{n+1}&=&\frac{\left({\bf F}_\gamma\right)^{n}}
{1+\Delta t n_{\rm H_0}^{n+1} c \sigma_\gamma}.
\end{eqnarray}
Injecting  these equations into  the rest  of the  system leads  to an
implicit system of 2 coupled  non--linear equations to solve for the 2
variables $x^{n+1}$  and $T^{n+1}$.  We  describe in the  Appendix the
technical solution we propose to  solve this problem. After this final
step, we  find the  new thermochemical state  $(x, T)^{n+1}$,  and the
corresponding new radiation state $(N_\gamma, {\bf F}_\gamma )^{n+1}$,
at the new time step $t+\Delta t$.

\section{Tests and Results}
In  their cosmological  radiative transfer  codes  comparison project,
\citet{2006MNRAS.371.1057I}  give a  set of  standard tests  that were
used to probe  the validity of ATON and  its implementation. First the
expansion of a Stromgren sphere is simulated, then the shadowing of an
I-front by  a dense clump is  discussed, and a first  attempt to model
the  propagation  of radiations  in  a  static  cosmological field  is
presented in the end.
 
\subsection{HII region expansion with a constant temperature}
\label{s:fixed}
\begin{figure*} 
\centering
\resizebox{0.8\columnwidth}{0.8\columnwidth}{\includegraphics{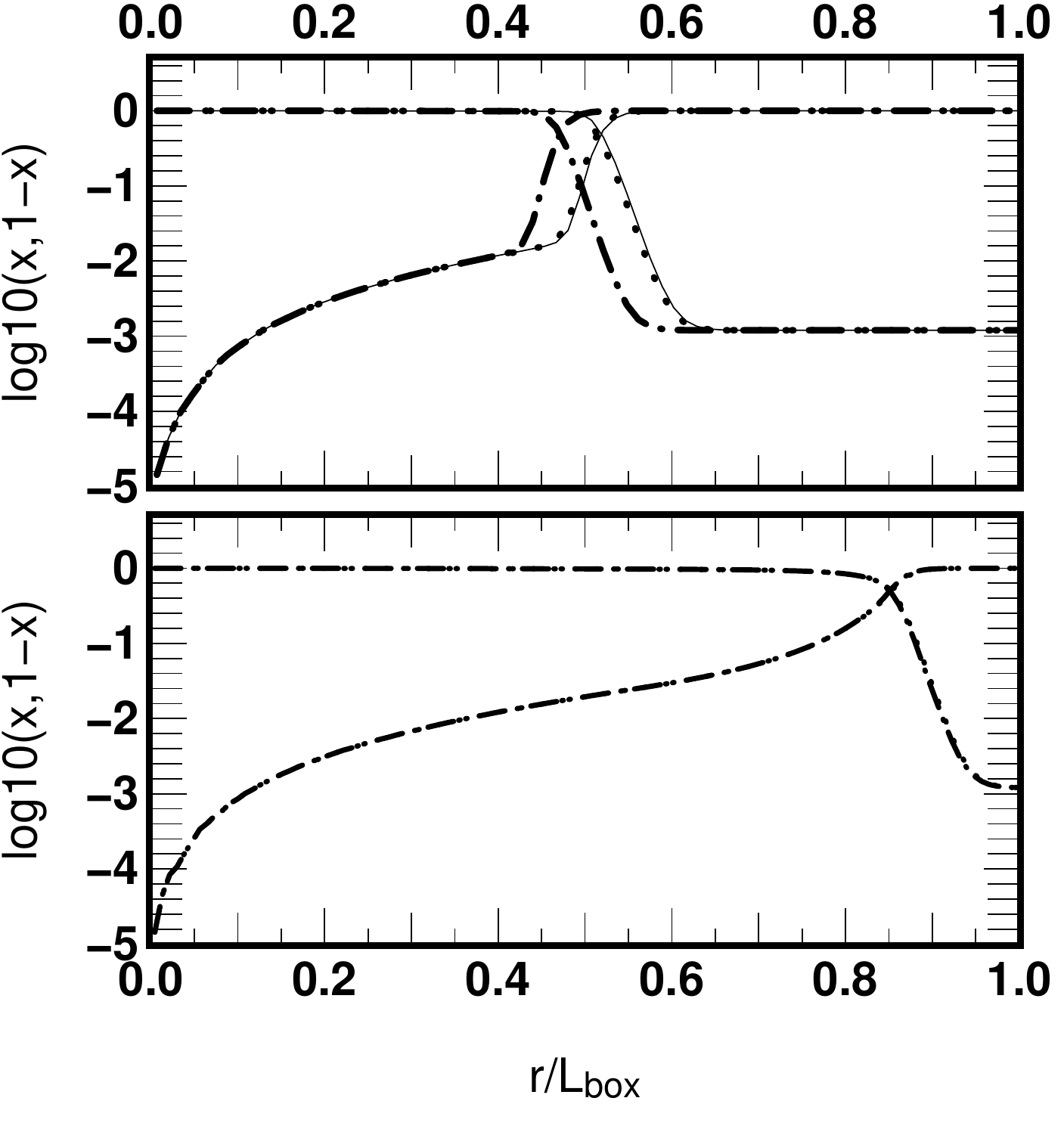}}
\resizebox{0.8\columnwidth}{0.8\columnwidth}{\includegraphics{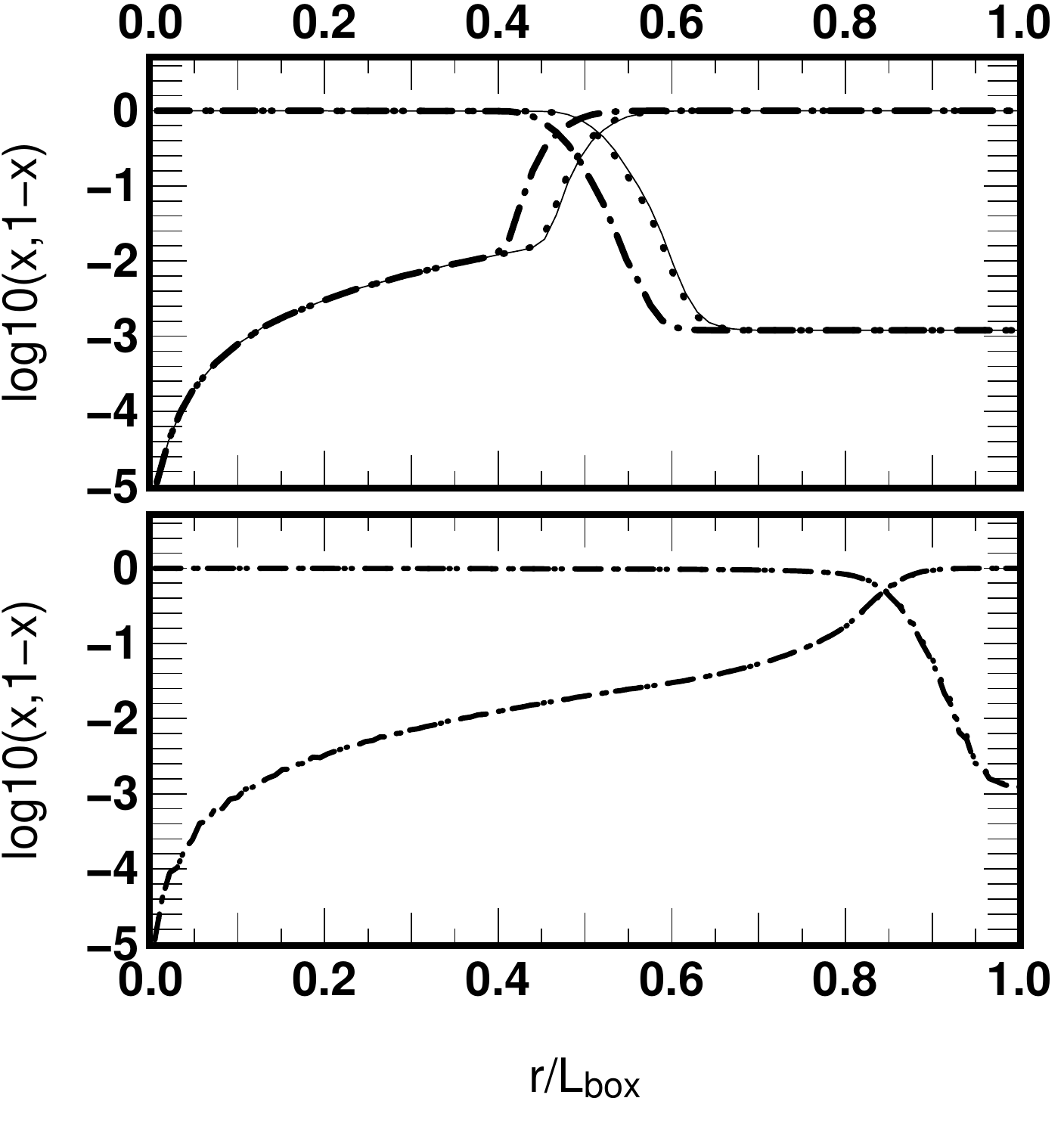}}
\caption{Stromgren sphere  test. GLF intercell  flux has been  used for
the left  panel's computations, HLL  intercell flux has been  used for
the  right  panel's  calculation.  The  lines stand  for  the  ionized
fraction profile and the  neutral fraction profile, computed at $t=35$
Myr (top  row) and  t= 500 Myr  (bottom row)  with a reduced  speed of
light $\tilde c =c  /1000$ (dot-dashed line), $\tilde c=c/100$ (dotted
line) and $\tilde c=c/100$ (plain line). Curves are superimposed at t= 500 Myrs.}
\label{f:stromprofile} 
\end{figure*}

\begin{figure*} 
\centering
\resizebox{0.8\columnwidth}{0.8\columnwidth}{\includegraphics{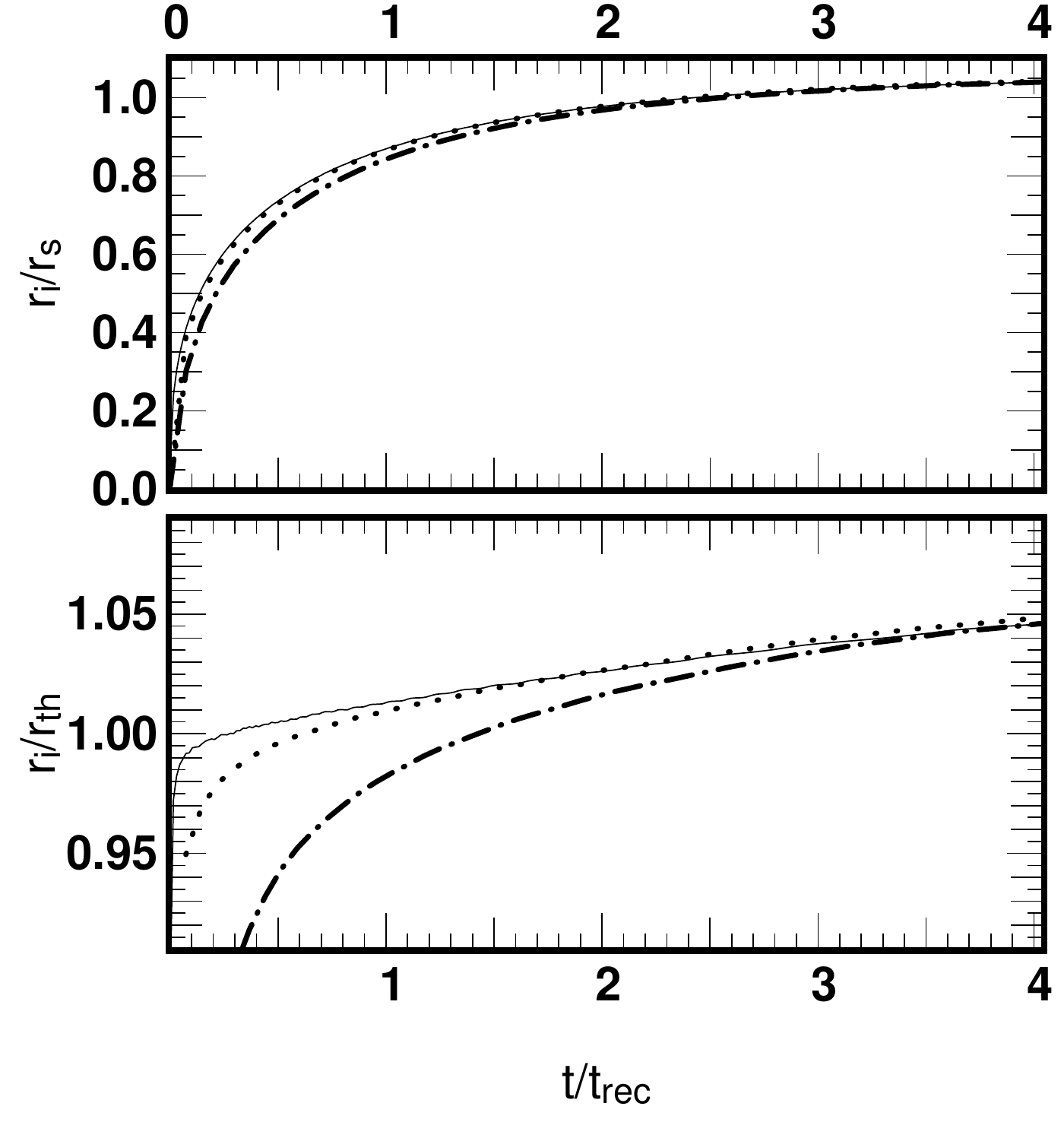}}
\resizebox{0.8\columnwidth}{0.8\columnwidth}{\includegraphics{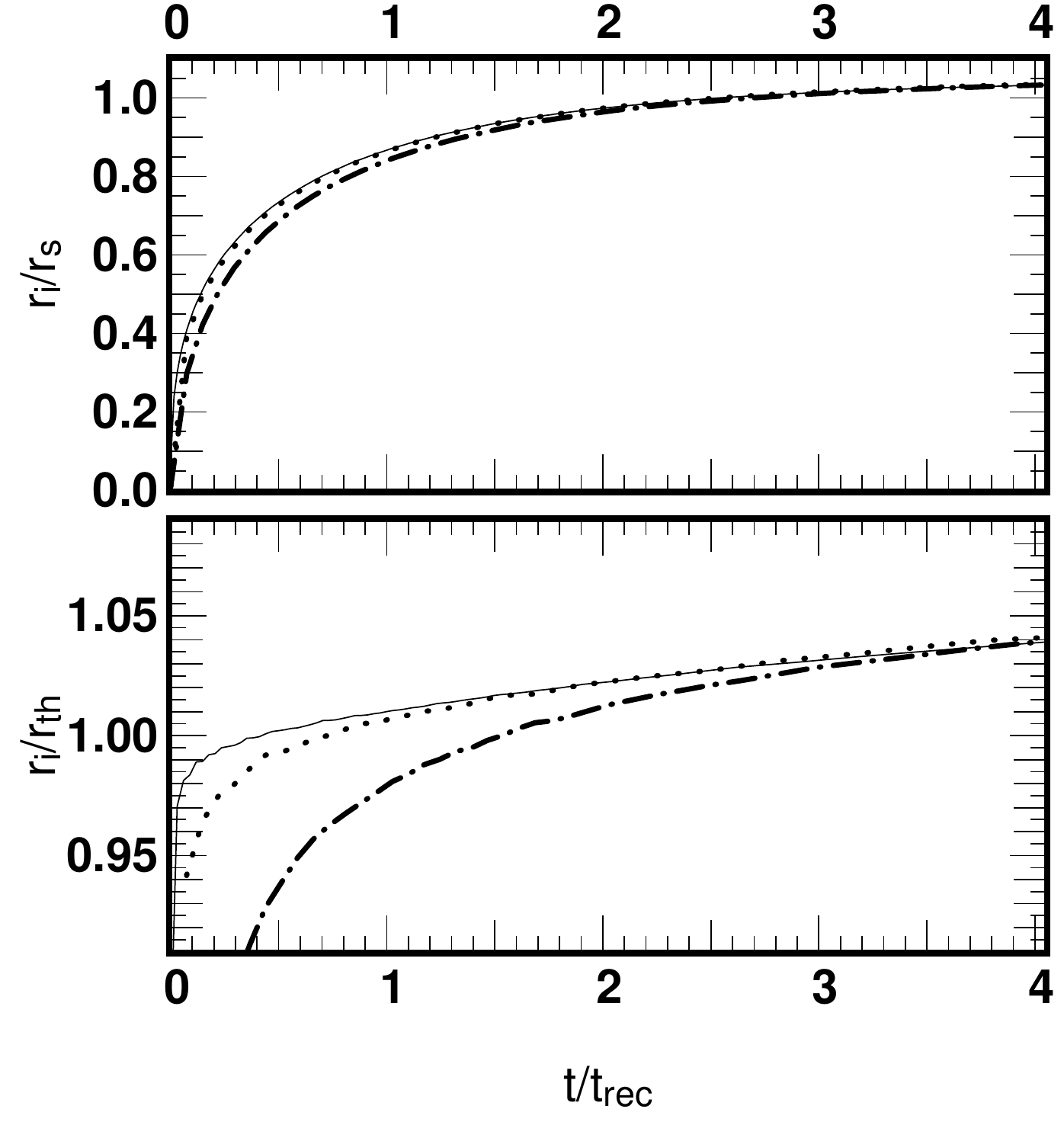}}
\caption{Stromgren sphere  test. The dotted (resp.  plain) lines stand
for  the  I-front  position  at  different  times  (in  units  of  the
recombination time), computed with a  reduced speed of light $\tilde c
=c /1000, c/100, c/10$ (resp. dot-dashed, dotted and plain lines). HLL
flux calculation is shown on  the right panels, GLF flux calculation on
the left  panels. \textit{Bottom rows:}  ratio of the  simulated front
position to  the analytic  value. \textit{Top rows:}  I-front position
evolution in  units of the box  size. The Stromgren radius  is 5.4 kpc
and the recombination time is 122.4 Myr.}
\label{f:stromevol} 
\end{figure*}

First, the classical situation where  a single source emits a ionizing
radiation that  propagates through  the surrounding neutral  medium is
presented. It results in the classical picture where an ionized sphere
, centered on the source,  expands at given rate. As the recombination
process starts to counter-balance the ionization, the front's position
slows down  and even stops,  achieving a stationary  regime.  Assuming
the source emits $\dot N^*_{\gamma}$  ionizing photons per unit time the
Stromgren radius, i.e. the stationary radius, is given by:
\begin{equation}
r_{S}=\left(\frac{3\dot N^*_{\gamma}}{4\pi\alpha_{B}(T)n_{H}^2}\right)^{1/3},
\end{equation}
where $\alpha_{B}(T)$  stands for  the case B  recombination rate  at a
temperature $T$ and $n_{H}$ is the surrounding gas number density. The
time evolution of the I-front's position is given by
\begin{equation}
r_{I}(t)=r_{S}(1-e^{-\frac{t}{t_{r}}})^{1/3},
\label{e:strompos}
\end{equation}
where  $t_{r}$  is  the  characteristic recombination  time  given  by
$1/t_{r}=\alpha_{B}(T)n_{H}$. From  Eq. \ref{e:strompos}, one  can see
that the expansion slows down for $t\sim t_{r}$.

The setting for  the numerical experiment is similar  to the one given
by \citet{2006MNRAS.371.1057I}. The source is located at the corner of
the simulated box and  emits $\dot N^*_{\gamma}=5\times 10^{48}$ photons
per  second.  The  surrounding   hydrogen  has  a  number  density  of
$n_{H}=10^{-3}\mathrm{cm}^{-3}$  with   an  initial  ionized  fraction
$x=1.2\times10^{-3}$  and its temperature  remains fixed,  with $T=10^4K$
. The  simulation is performed  on $64^3$ grid,  with a $6.6$  kpc box
size.  Reflexive  boundary  conditions  are  assumed.  The  source  is
switched    on    at   t=0.    The    results    are   presented    in
Figs. \ref{f:stromprofile} and \ref{f:stromevol}.

Fig.  \ref{f:stromprofile}  shows  the  profiles of  the  ionized  and
neutral fraction  computed with the  two different types  of intercell
flux at  t=30 Myr and t=500Myr.  Clearly the two methods  agree for an
effective speed  of light  that varies  from $\tilde c  = 0.001  c$ to
$\tilde  c=c$. For  instance,  both methods  return an  underestimated
I-Front  position   as  $\tilde  c$  is  smaller   than  the  physical
propagation speed of  the front. Conversely, when $\tilde  c$ is large
enough   (typically  $\tilde   c  \ge   0.01  c$),   the   two  fluxes
approximations agree and achieve convergence regarding the result. The
same calculation  is performed in  \citet{2006MNRAS.371.1057I} and the
results  shown in  Fig. \ref{f:stromprofile}  are consistent  with the
profiles  found  by the  codes  that  took  part to  their  comparison
project.

In  \Fig{stromevol},  the  I-front's   position  is  compared  to  the
theoretical calculation.  From the  top panels of  \Fig{stromevol}, it
clearly appears  that the current code accurately  reproduces the fast
expansion of  the ionized bubble at  early times and the  slow down as
the I-front  position gets  closer to the  Stromgren radius.  At later
times, a  stationary state is  achieved and the I-front  stops. Again,
the  GLF and HLL  calculation agrees  and return  a final  radius $4\%$
larger  than  the  theoretical  Stromgren radius.  The  comparison  to
\citet{2006MNRAS.371.1057I} shows that it is consistent with the other
types of  calculations which tends  to overestimate the  final radius'
value.  The bottom panels  show the  comparison between  the simulated
I-Front position  and the theoretical  one at each time  step. Clearly
the two types of intercell  fluxes (GLF and HLL) return similar results
and  both  present  a  drift   toward  greater  values  of  the  front
position. For t=500 Myr, the discrepancy is $4\%$. Interestingly, this
drift is  similar for all  values of $\tilde  c$ and the  results only
differ by the  time required for the profile  to 'converge' toward the
drift. For instance, the  $\tilde c=0.001c$ calculation requires $\sim
t_{\mathrm{rec}}$   to  catch  up   the  main   drift,  while   it  is
instantaneous   for   $\tilde    c=0.1c$.   The   results   found   in
\citet{2006MNRAS.371.1057I} shows a similar behavior.

\subsection{HII region expansion with a variable temperature}
In the second test, the HII region expansion is investigated while the
gas   temperature   is  allowed   to   vary   as   a  consequence   of
photoheating. The  numerical setup is  identical to the  previous test
with two  exceptions. First, the source  is now a $10^5$  K black body
source while  the flux of ionizing photons  remains unchanged. Second,
the initial  gas temperature is  set to $100$  K and its  evolution is
modeled according to the \Eq{temp}.   The source is turned on at $t=0$
and the  simulation is  run over one  recombination time  with $\tilde
c=0.1 c$.

\begin{figure*} 
\centering
\resizebox{0.8\columnwidth}{0.8\columnwidth}{\includegraphics{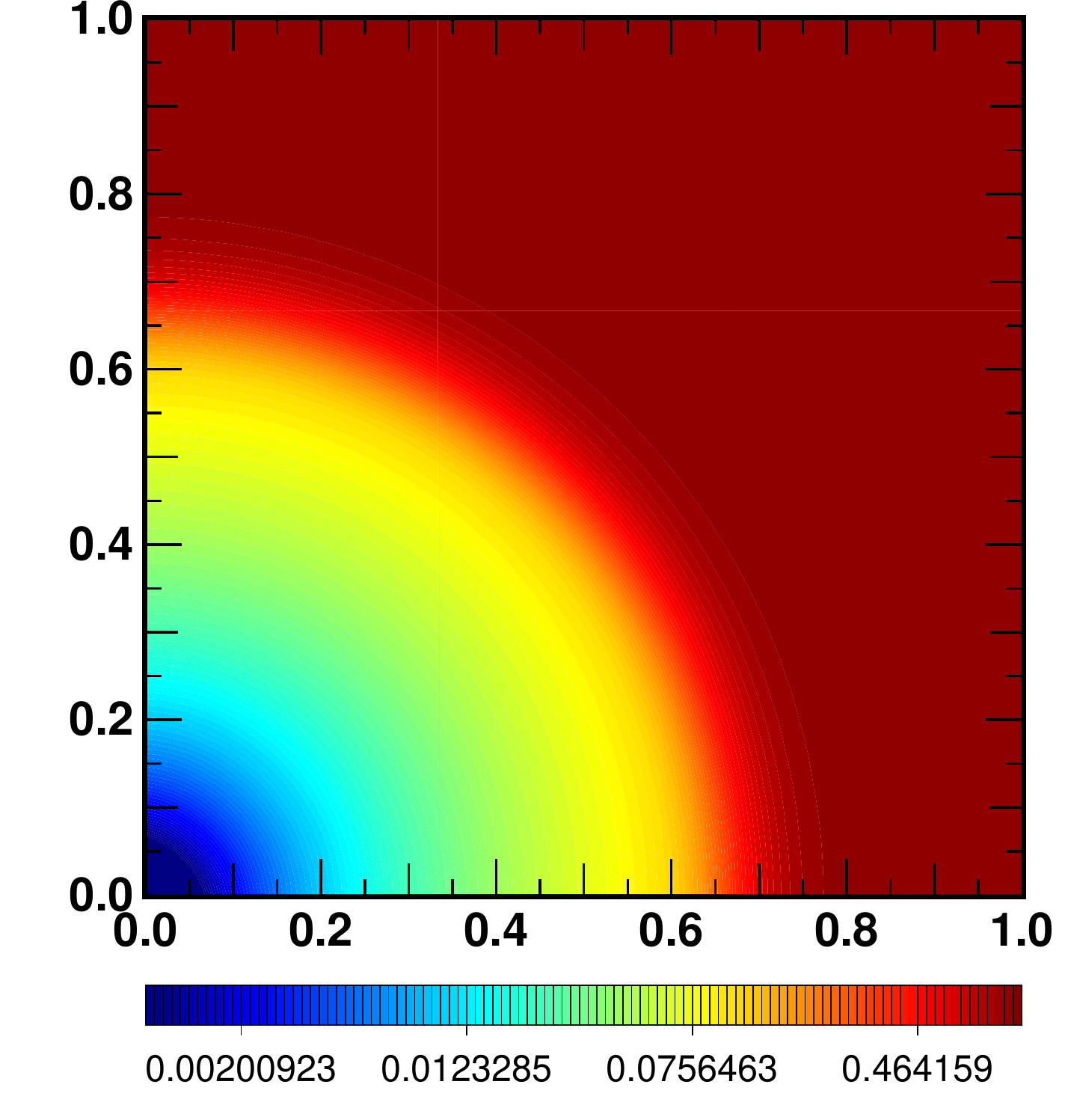}}
\resizebox{0.8\columnwidth}{0.8\columnwidth}{\includegraphics{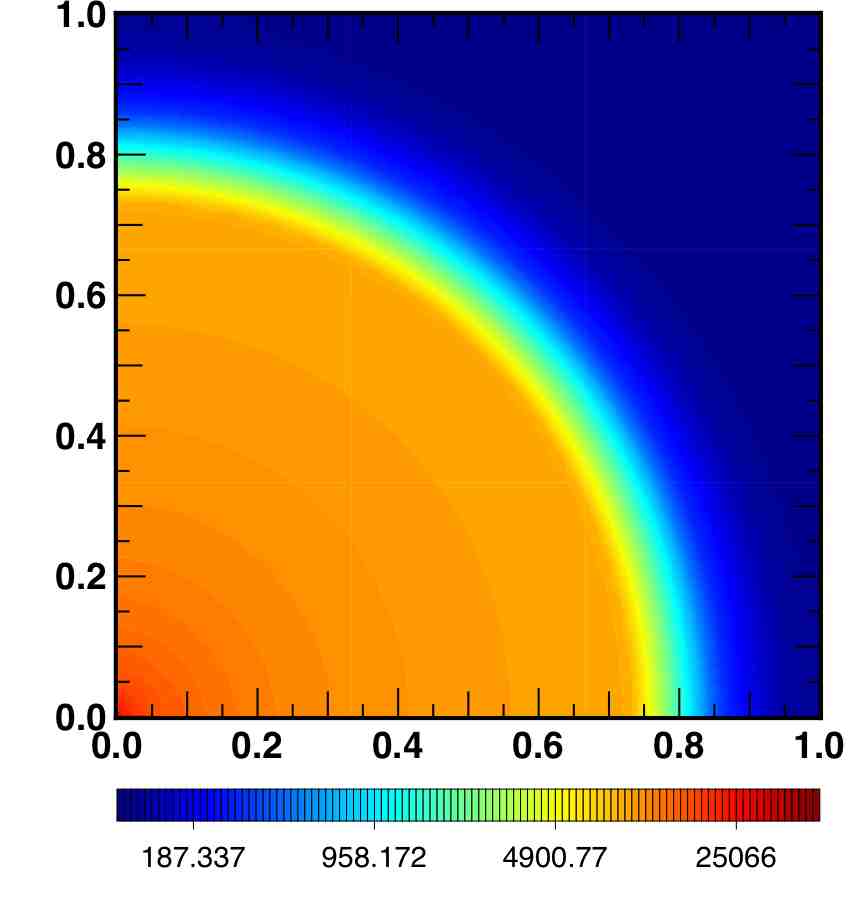}}
\caption{Stromgren      sphere     test      with      non     uniform
temperature.  \textit{Left~:}  Neutral  fraction  map  in  the  source
plane.  \textit{Right~:} temperature  (in  Kelvin) map  in the  source
plane.  These maps  were computed  100  Myr after  the central  source
(located here  in the  bottom left corner)  has been switched  on. The
experimental  setting is  similar  to the  constant temperature  test,
described in section \ref{s:fixed}.}
\label{f:maptemp} 
\end{figure*}

\begin{figure*} 
\centering
\resizebox{0.9\columnwidth}{0.8\columnwidth}{\includegraphics{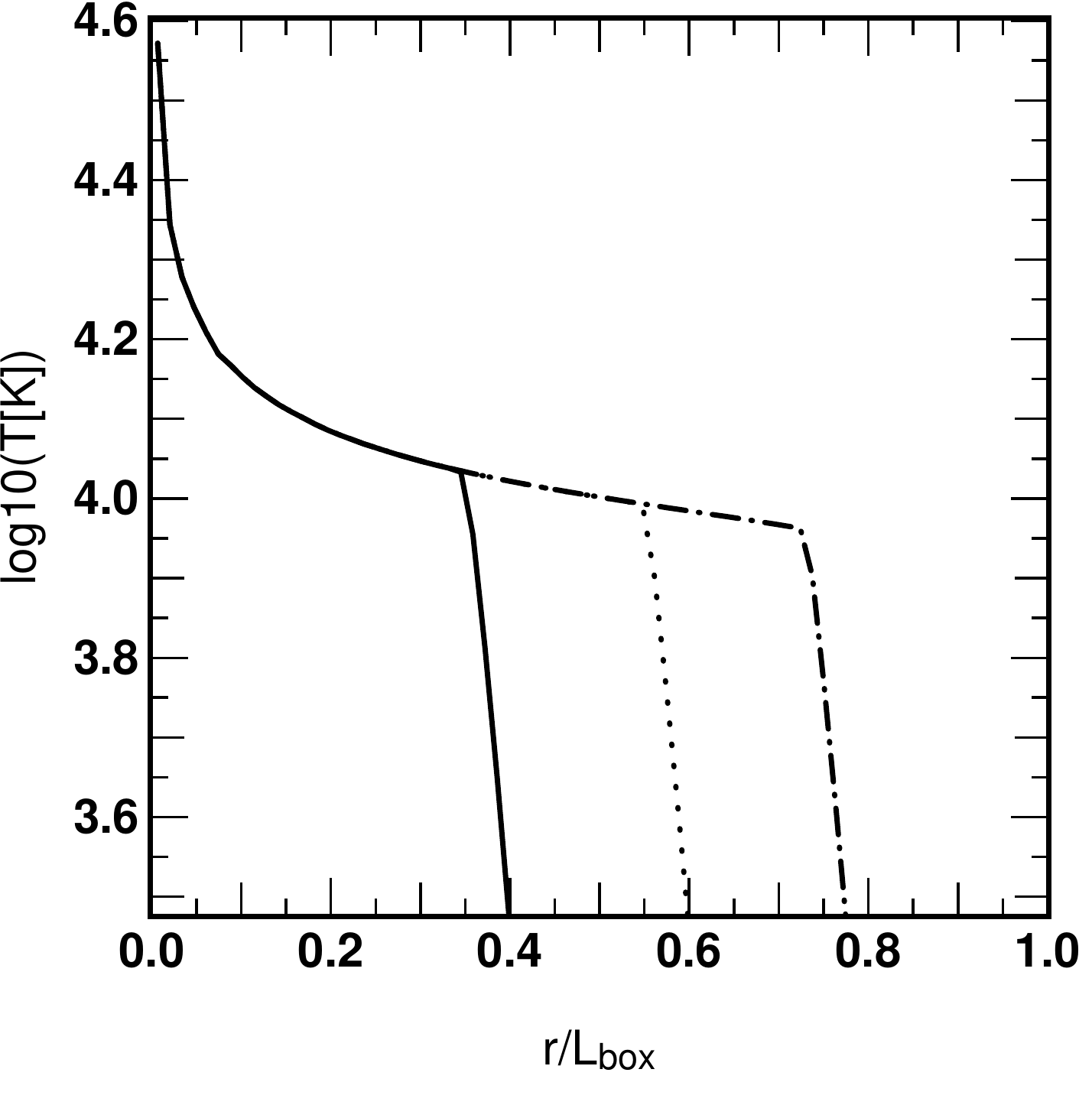}}
\resizebox{0.9\columnwidth}{0.8\columnwidth}{\includegraphics{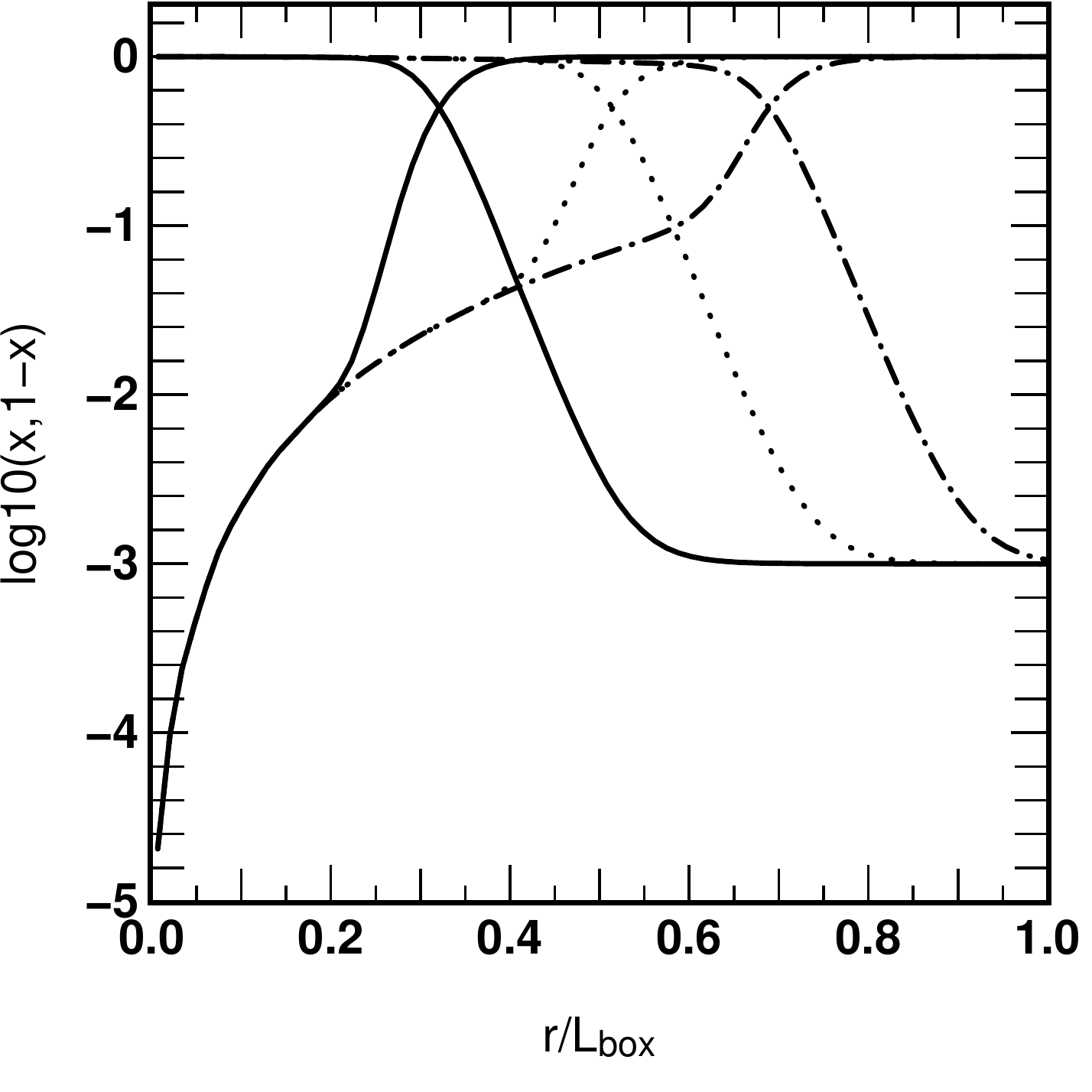}}
\caption{Stromgren sphere test with  non uniform temperature. The left
panel  shows the  average temperature  profile while  the  right panel
shows the  the average ionization and neutral  fraction profile. These
profiles were computed 10(plain),  35 (dotted) and 100 (dashed-dotted)
Myrs after the  central source has been switched  on. The experimental
setting  is similar  to the  constant temperature  test,  described in
section \ref{s:fixed}.}
\label{f:stromtemp} 
\end{figure*}

\Fig{maptemp} presents the temperature and neutral fraction maps after
100  Myrs  and \Fig{stromtemp}  presents  the  time  evolution of  the
temperature and  ionized fraction profile. A  quick comparison between
the constant temperature experiment and the current one indicates that
the results  qualitatively agrees. For instance, taking  the t=35 Myrs
profile as a reference, the  transition region is in both case located
at a  radius equals to 0.45  , in units  of the box size.  However the
transition  operates  on a  larger  range  of  radii compared  to  the
isothermal  case, the  background neutral  fraction being  achieved at
r=0.8  box length  at  t=35 Myrs  while  this region  does not  extend
further than r=0.65 in the  constant temperature case. This relates to
the non-monochromatic  source used in  the current case~:  the average
photon-energy   is   larger  than   previously,   hence  the   average
cross-section is  diminished and this  leads to a  smoother transition
from the fully ionized to  the neutral regime. The temperature profile
shows a steady evolution of the limit between the hot medium (T$\ge 10
000$K) and the  cold neutral medium. The transition  between these two
regimes  is  much  sharper  than  observed for  the  neutral  fraction
profile. Furthermore the inner temperature profile is practically flat
(except close to  the center) while the neutral  fraction values spans
over a  few orders of magnitude. It  can also be observed  on the maps
where no dynamics can be seen on the temperature maps while a gradient
in the neutral fraction is clearly visible.

Comparisons with \citet{2006MNRAS.371.1057I}  results show that ATON's
calculations  are   consistent  with  other   codes,  however  
%several
%differences can  already be  noted. 
%First the  equilibrium temperature
%obtained in  the ionized region  is slightly ($\sim10 \%$)  lower than
%most of the \citet{2006MNRAS.371.1057I} computations.
%This is easily explained by the speed-up procedure applied here for the heating-cooling computation~: since the temperature's calculation is stopped at a certain level of convergence, we do not inject all the energy which is supposed to photo-heat the gas. 
the  transition region's  profile between ionized  and neutral
gas is sharper in the  current calculations than returned by the other
codes.  
%This could be explained  by the cooler temperature which leads
%to higher recombination rates. 
It could be related to the current
spectral treatment, where a single photon-population (having an energy
representative  of  the source  spectrum)  is  taken  in account,  the
transition's profile reflecting its typical energy. Conversely, a more
complete  treatment  of  spectral  hardening is  expected  to  produce
smoother profiles,  as high-energy photons have  larger mean free-path
and travel further in the neutral hydrogen.

\subsection{Shadowing by a dense clump}
The third  test investigates the  code's ability to deal  with density
clumps along the photon's path.  Such clumps are expected to slow down
the I-fronts  propagation and create  regions which are  shielded from
the  radiations  due  to  the  shadow  trailing  behind  high  density
regions.  Such  situations are  likely  to  happen  in a  cosmological
context  where shielding  is provided  by e.g.  gaseous  structures or
filaments.  The setting  suggested  by \citet{2006MNRAS.371.1057I}  is
adopted: a  6.6 kpc box is considered  with a dense and  cold clump on
the  radiation's  path.  The  background  gas  density  is  $\rho=200$
m$^{-3}$ with  a initial temperature  $T=8000$ K. The clump  is sphere
located at (5, 3.3, 3.3) kpc with a radius $r=0.8$ kpc. Its density is
$40  000$ m$^{-3}$  and its  temperature  is $T=80$  K.  A  stationary
photon flux $\Phi=10^{10}$m$^{-3}$s$^{-1}$ photons is ignited at $t=0$
with  a  typical  energy   corresponding  to  a  $10^5$  K  black-body
source. Calculations  were performed with the HLL  intercell flux with
$\tilde c=c$.

\begin{figure} 
\centering
\resizebox{0.8\columnwidth}{0.8\columnwidth}{\includegraphics{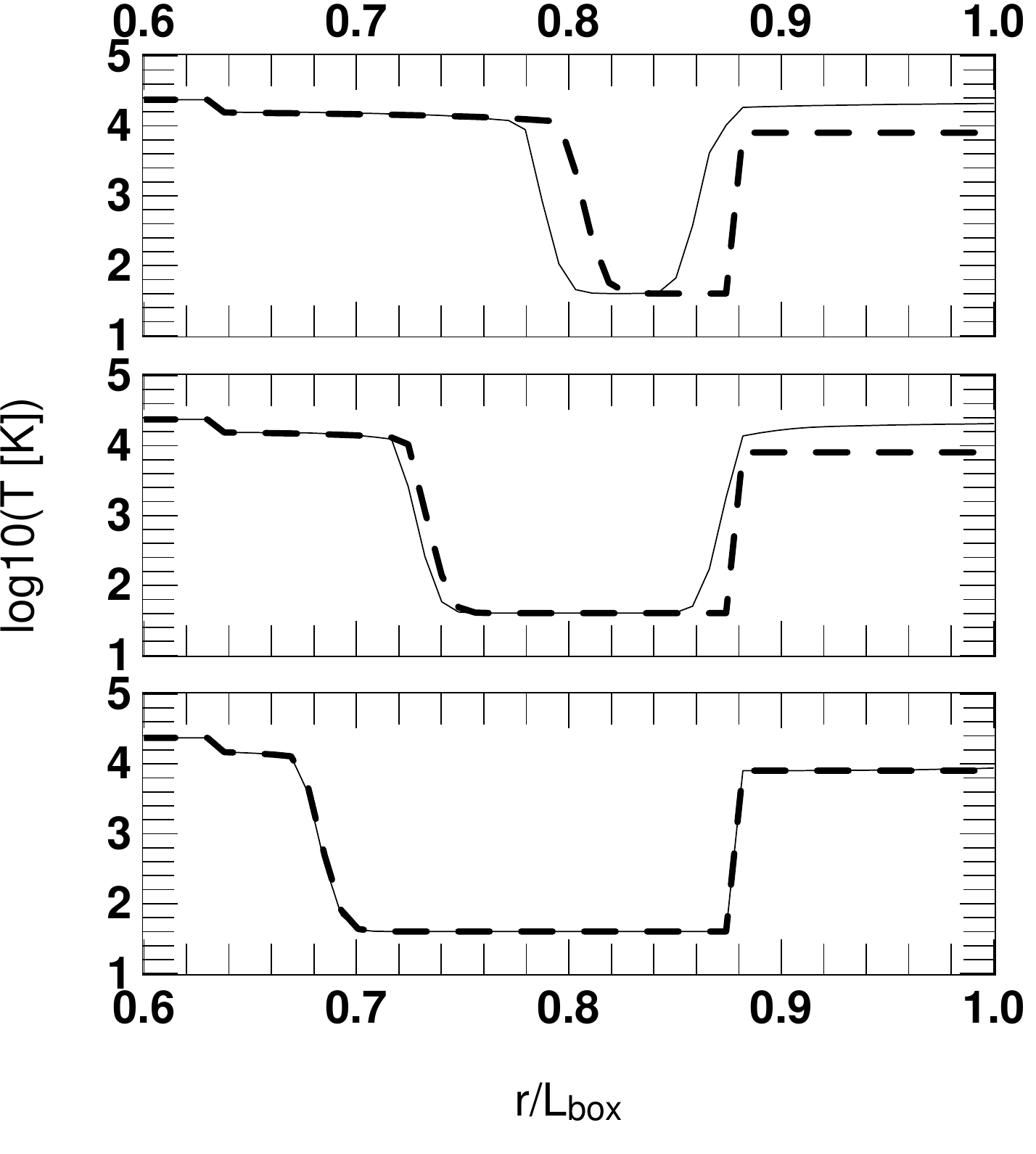}}
\resizebox{0.8\columnwidth}{0.8\columnwidth}{\includegraphics{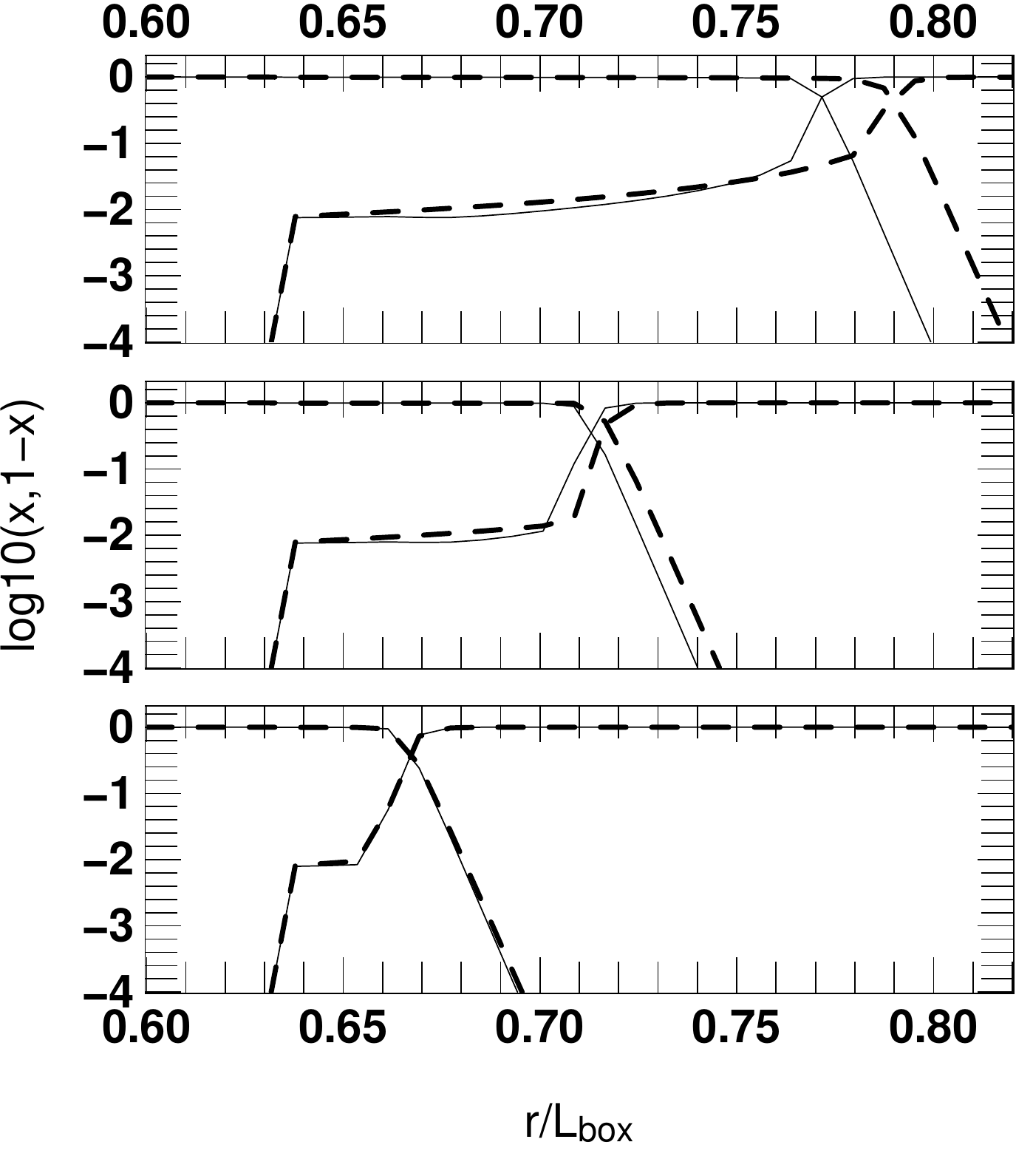}}
\caption{Shadowing  by a dense  clump using  HLL intercell  flux. Top:
Temperature  profile along  the middle  section of  the  box.  Bottom:
Neutral and ionized fraction  profile along the same section. Profiles
were computed  at t= 1 (bottom row),  3 (middle row) and  10 (top row)
Myrs. The radiation  comes from the left side of  the maps. Thin solid
lines stand  for the computation  using the 'regular' M1  scheme while
thick   dashed   lines   stand   for   the   computation   using   our
``ray--tracing'' scheme.}
\label{f:mapclumpx} 
\end{figure}

\begin{figure} 
\centering
\resizebox{0.8\columnwidth}{0.8\columnwidth}{\includegraphics{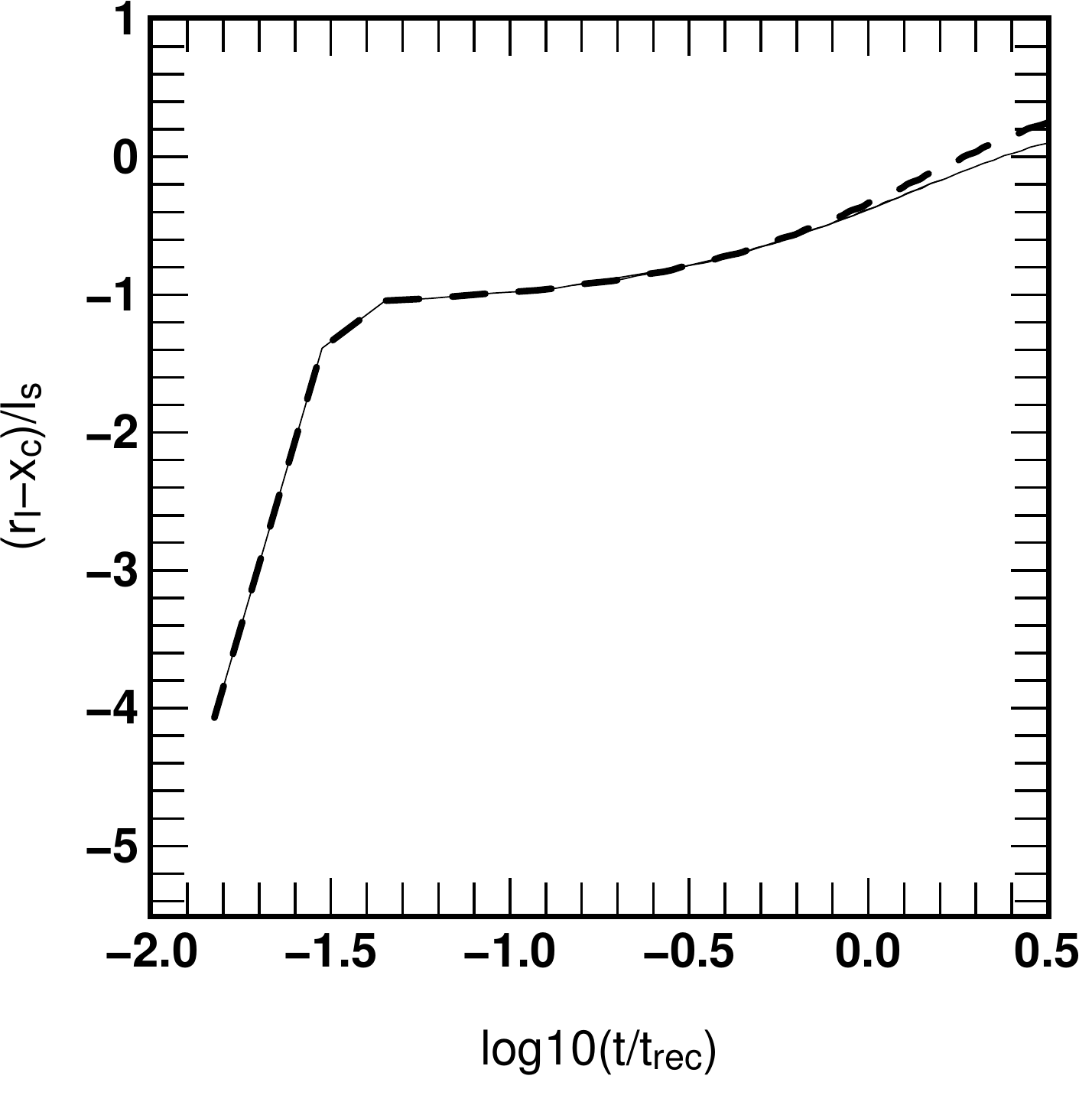}}
\caption{Shadowing  by  a  dense  clump.  The I-front  position  as  a
function of time  for the regular ATON's scheme  (thin solid line) and
``ray--tracing'' scheme (thick dashed line). The position of the I-front, relative to the clump center is given  in units of  the Stromgren length  defined as $\ell_{S}=F/\rho^2\alpha_{B}$. }
\label{f:clumpipos} 
\end{figure}

\begin{figure} 
\centering
\resizebox{0.8\columnwidth}{0.8\columnwidth}{\includegraphics{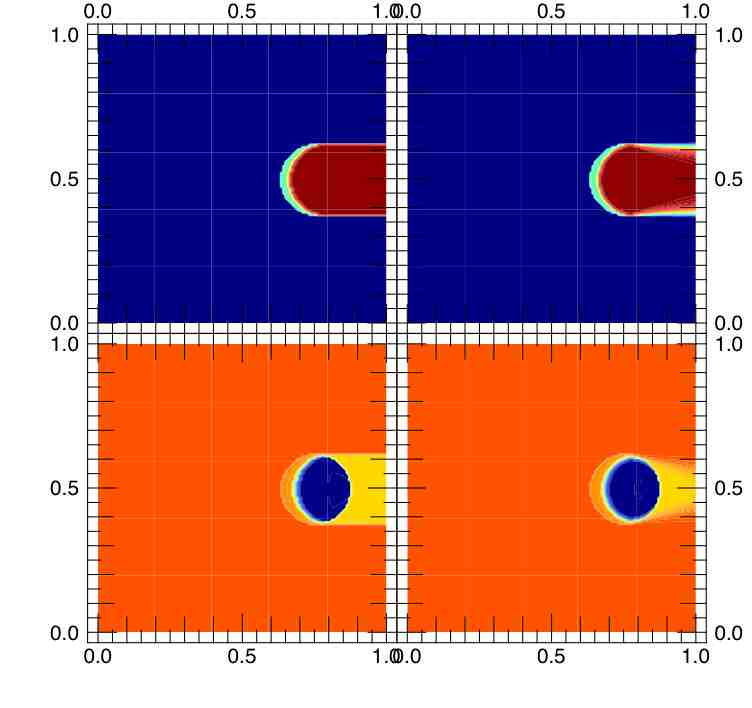}}
\resizebox{0.8\columnwidth}{0.8\columnwidth}{\includegraphics{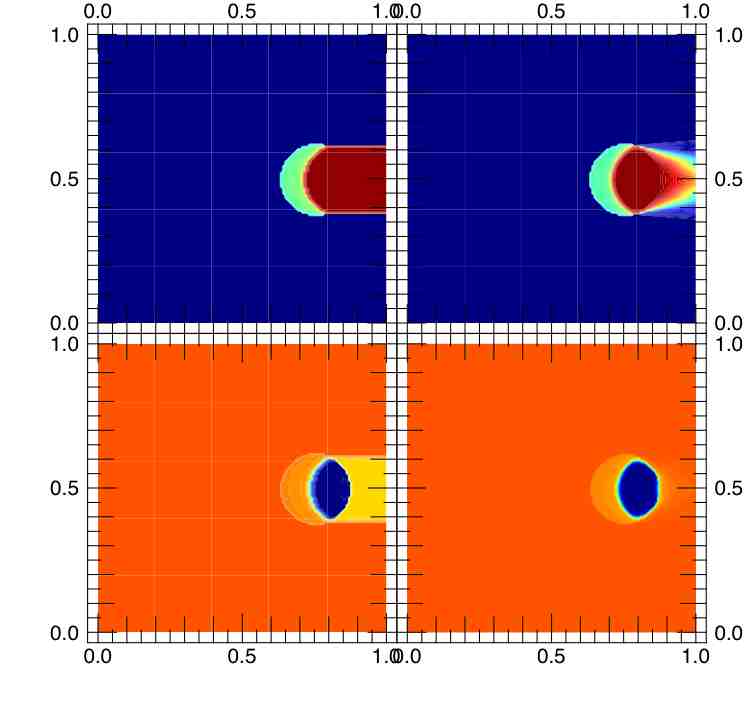}}
\caption{Shadowing  by a dense  clump using  HLL intercell  flux. Top:
neutral fraction (top row) and  temperature map (bottom row) along the
middle  section   of  the  box,  measured  at   t=1Myr.  Bottom:  same
measurements at t= 3 Myrs.  Left columns stand for computations with a
``ray--tracing''  scheme, while right columns stand  for the 'regular'
M1 scheme. }
\label{f:mapclumpx2} 
\end{figure}

\begin{figure} 
\centering
\resizebox{0.8\columnwidth}{0.8\columnwidth}{\includegraphics{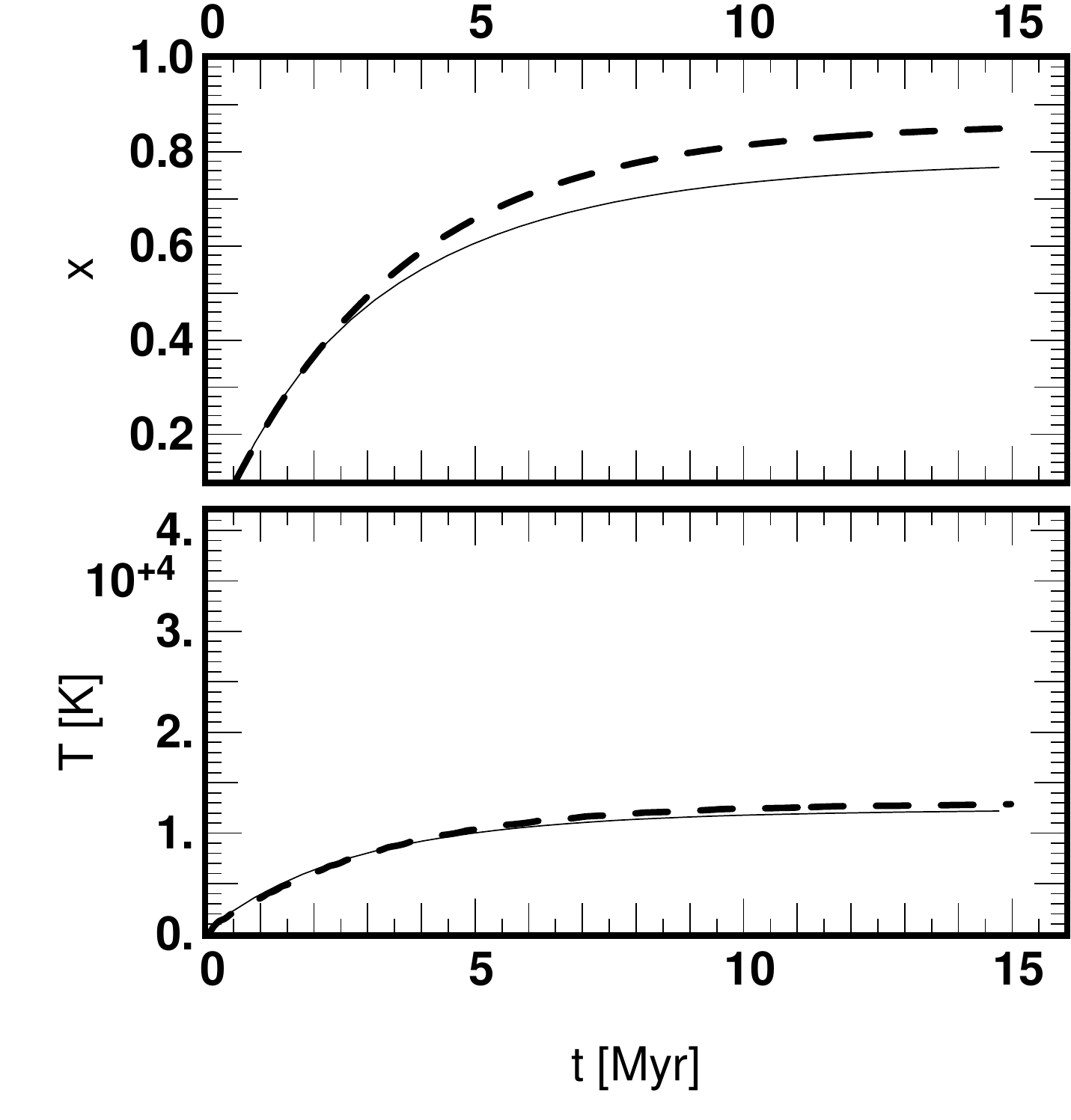}}
\caption{Shadowing by a dense clump. The time evolution of the average
ionized fraction (top) and the average temperature (bottom) within the
clump  for  the  regular  ATON's  scheme (thin  solid  line)  and  the
``ray--tracing'' scheme (thick dashed line). }
\label{f:clumpevol} 
\end{figure}

Neutral  fraction  and temperature  maps  and  profiles  are given  in
\Fig{mapclumpx2} and \Fig{mapclumpx}. Clearly  shadows are created in the  clump trail: the
gas remains neutral behind the clump and its temperature remains lower
than  the surroundings  due to  the lack  of heating  photons  in this
region.  The  neutral fraction is almost zero outside the
clump due to the incoming flux of photons ad rises to a constant level
around $x\sim0.01$ on the `enlightened'  side of the clump as light penetrates the over-density. 
%At some distance from the illuminated border of the clump
%the  medium  becomes fully  neutral.  
As illustrated by \Fig{clumpipos}, the  I-front
propagates deeper into the clump and after a
fast propagation in the optically  thin medium, the front almost stops
and  progress  slowly  trough   the  dense  medium.  
%Comparisons  with
%\citet{2006MNRAS.371.1057I} indicates  that ATON reproduces accurately
%this  process. 
On  the other  hand, the  temperature profiles  shows a
similar behavior  with a sharp transition region  from $\sim 10500-11
000$ K to  $80$ K, the initial clump's  temperature. In the 'trailing'
region,   temperature  is   initially  lower   than  in   the  leading
region. However, it  is clear from the maps  in \Fig{mapclumpx2} that
the limits of the trailing shadow are not parallel to the direction of
the incident flux, and a diffusion cone appears.  As time advances, the temperature in  the initially shielded region rises, until it  reaches a  level close  to the  one observed  in the exposed gas.  
Finally, \Fig{clumpevol} describes  the time  evolution of the  average ionized
fraction and  temperature inside the  clump. Both curves shows  a fast
increase  during  the first  million  years  followed  by a  shallower
evolution  toward $x=0.8$  and $T=11000$  K.  Up to  t=10 Myrs,  these
evolutions are  in a quantitative  agreement to the ones  presented in
\citet{2006MNRAS.371.1057I}, which  indicates that the  current scheme
captures the overall picture of  the I-front trapping, even though the
lack of  multi-frequency treatment leads to some  discrepancies in the
detailed  description of the  process. 

From  a   physical point-of-view, a certain   level   of  diffusivity   is   expected~: atoms recombine and radiate in an isotropic fashion. In
the  current scheme,  isotropy is  taken in  account by  the spherical
component of  the Eddington's tensor and its  relative contribution is
set by the  ratio of the local  flux by the energy (  the reduced flux
$f$). From the  maps in \Fig{mapclumpx}, the clear  conic shape of the
trailing  shadow appears  as a  manifestation of this isotropy. 
However, ATON's predicts  a larger  shadow's extent compared  to experiments in \citet{2006MNRAS.371.1057I}  where it remains parallel and clear  cut~: this discrepancy is likely to be related to the lack of modelisation of the recombination's isotropy.  
As stated in  the first section,  ATON's can mimic
the  behavior  of  a  ray-tracing   code  by  adding  an  extra  {\it
artificial}  term in the  flux equation  (see \Eq{rt1}  and \Eq{rt2}).
The computations  using this  scheme are significantly  different from
the  one using  the regular  scheme and are in good agreement with results shown in \citet{2006MNRAS.371.1057I} . In  \Fig{mapclumpx2},  it clearly
appears that the  'ray--tracing' scheme is  much more 'conservative'
in  terms of  flux  geometry: the  shadows  are clear  cut behind  the
clump.  It is  particularly  striking in  the  temperature maps  where
basically no  shadow persists  after a given  duration in  the regular
calculation while it  remains and exhibits sharp edges  in the current
enhanced  calculation. The  same effect  can be  noted in  the neutral
fraction maps where  the cone-shaped shadow is replaced  by a straight
cylindrical one. It  persists over the 15 Myrs  of the calculation. In
terms   of   profiles,   it   can  be   seen   from   \Fig{mapclumpx},
\Fig{clumpevol}  and \Fig{clumpipos} that  the I-front  propagation is
affected too and  their progression through the clump  is faster. 
%This
%is not  completely surprising since  the structure of the  newly added
%term in the flux equation  implies that the radiation's geometry  tends to remain
%somewhat  constant:  
The  radiation's  flux  is  more  'rigid'  and  the
influence  of  diffusion  is   lowered~:  the  ionization  of  a
propagating front is more  efficient especially in dense regions where
I-front can  be significantly  slowed down. Also, the
smaller  diffusivity  keeps the  initial temperature  profile in the shielded region unchanged. Let us emphasize that these results are obtained using an artificial flux term and should be considered with caution since diffusivity must appear at some point. As shown hereafter,
 the differences remain small in realistic cosmological situations but still~: in the current setting clear cut shadows are obtained by suppressing the isotropic recombination and as shown in the results obtained using the regular ATON's scheme, results can be significantly different in certain situations.

The other differences  with  the  calculations presented  in   \citet{2006MNRAS.371.1057I} can be explained by the lack of multi-frequency treatment.  Because the  scheme do  not take  in  account the
effect  of high  frequency photons  with large  mean free-path, no
preheating  on large  distance  is being  modeled.   With the  current
single-energy  treatment,  the effect  of  these  photons  tend to  be
underestimated and all  processes (ionization and photo-heating) occur
on small scales, leading to the observed sharp profiles. In \Fig{mapclumpx} ,  the  ionization
fraction transition  from ionized background to neutral  clump is much
sharper  in ATON's case   and  no  steady  decrease   can  be
observed. Other codes do not  present a sharp drop in ionized fraction
and a smoother  transition exists between the neutral  fraction of the
irradiated side  to the shielded side.  A similar
discrepancy  exists in  the  temperature profile,  where  the drop  in
temperature  occurs  on smaller  distances.  The  average ionized
fraction within the clump is also a few percent smaller than found in other
calculations at late times. In  other words, the  progression of the  I-front within
the clump  appears to be  slightly slower in ATON's  calculations, the
trapping  is  more efficient.  It  results  from the  lack  of
preheating and ionization  by high energy photons. In  the prospect of
cosmological studies, where sources are  likely to be embedded in high
density  regions,  the  radiation  is  therefore expected  to  take  a
slightly  larger amount  of  time to  escape  from the  haloes in  the
current scheme compared to other calculations.

 %To conclude, it is not clear  yet how excessive is the diffusivity of
 %the M1  scheme~: it must  be physically present, however  the results
 %are  quite different than  the one  obtained by  other codes.  On the
 %other side, the results of other calculations are recovered by adding
 %a supplementary yet  ad hoc source term in the flux equation. At  the current stage, we
 %aim  at performing all  calculations using  both schemes,  keeping in
 %mind this intrinsic diffusivity of the scheme but 

\subsection{Static Cosmological Field}
\begin{figure} 
\centering
\resizebox{0.8\columnwidth}{0.8\columnwidth}{\includegraphics{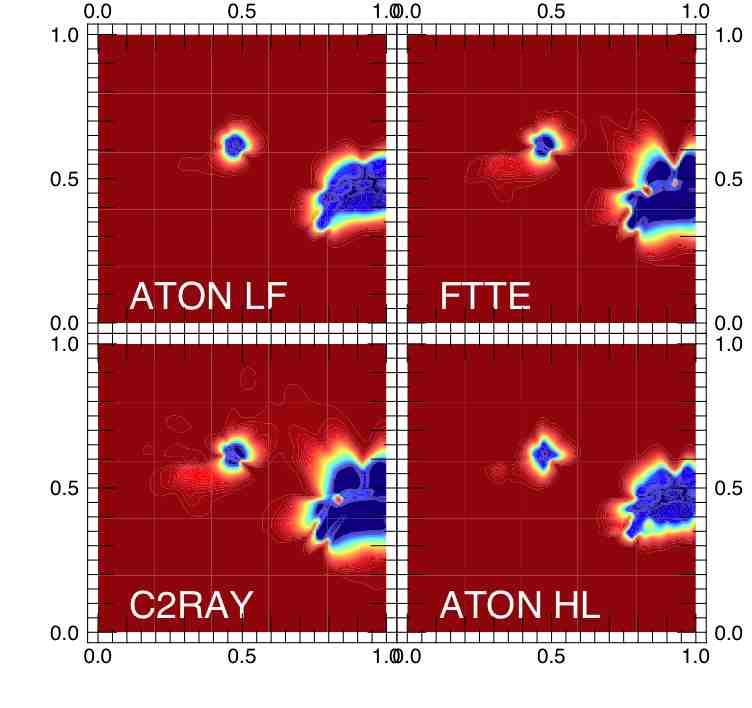}}
\resizebox{0.8\columnwidth}{0.8\columnwidth}{\includegraphics{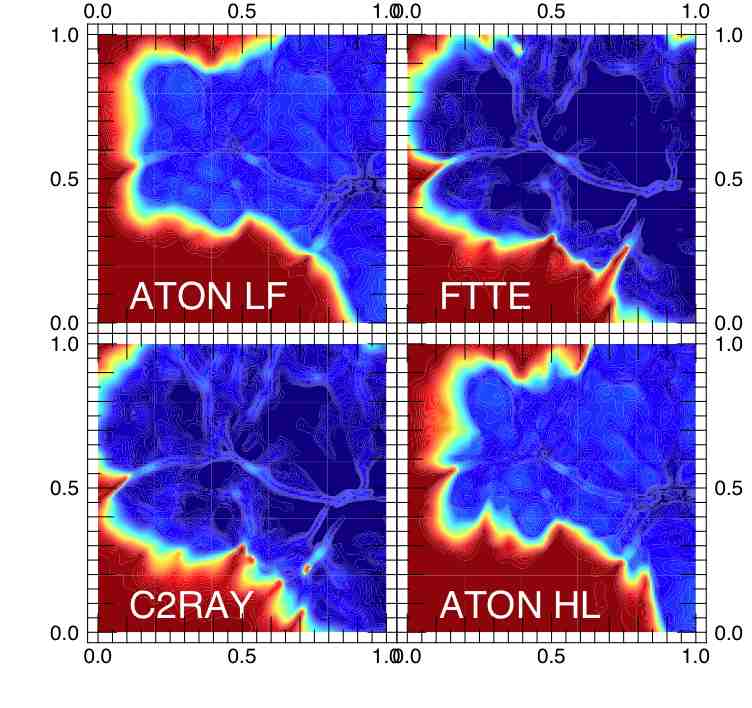}}
\caption{Static cosmological field test.  Maps of the neutral fraction
computed  at 0.05  Myr  (\textit{Top}) and  0.4 Myr  (\textit{bottom})
after the sources were ignited.  the four calculations were made using
ATON-GLF, ATON-HLL, C2RAY and FTTE.}
\label{f:mapcosmox} 
\end{figure}

\begin{figure} 
\centering
\resizebox{0.8\columnwidth}{0.8\columnwidth}{\includegraphics{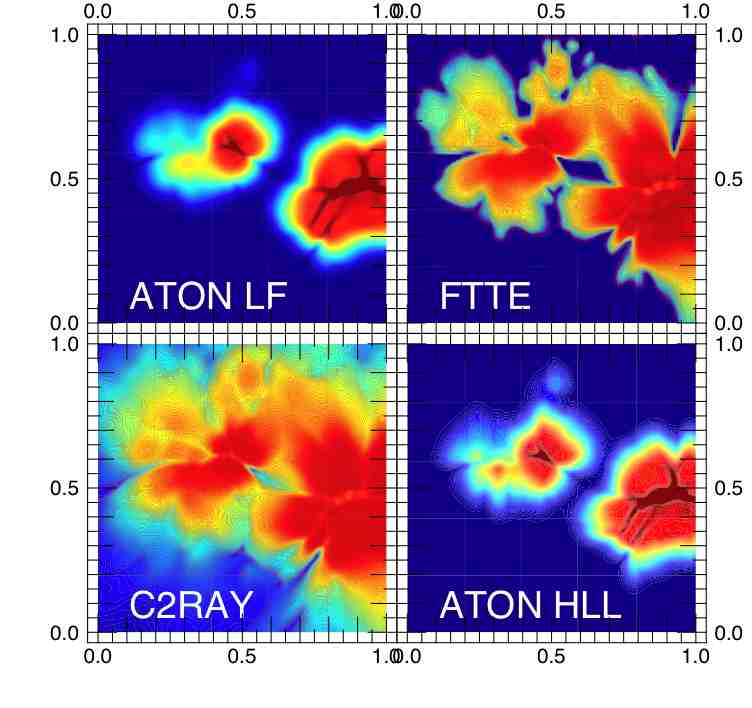}}
\resizebox{0.8\columnwidth}{0.8\columnwidth}{\includegraphics{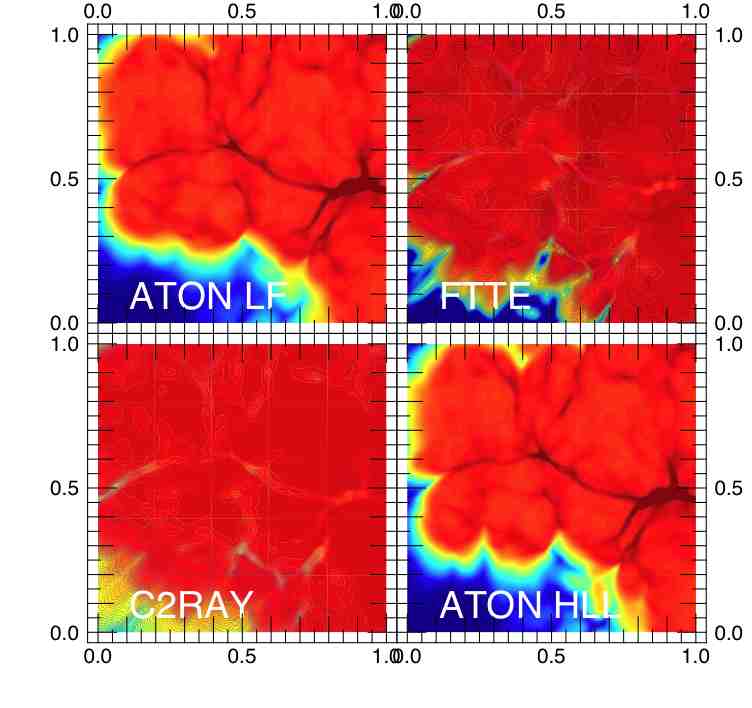}}
\caption{Static  cosmological  field  test.  Maps of  the  temperature
computed  at 0.05  Myr  (\textit{Top}) and  0.4 Myr  (\textit{bottom})
after the sources were ignited.  the four calculations were made using
ATON-GLF, ATON-HLL, C2RAY and FTTE.}
\label{f:mapcosmox2} 
\end{figure}

\begin{figure} 
\centering
\resizebox{0.8\columnwidth}{0.8\columnwidth}{\includegraphics{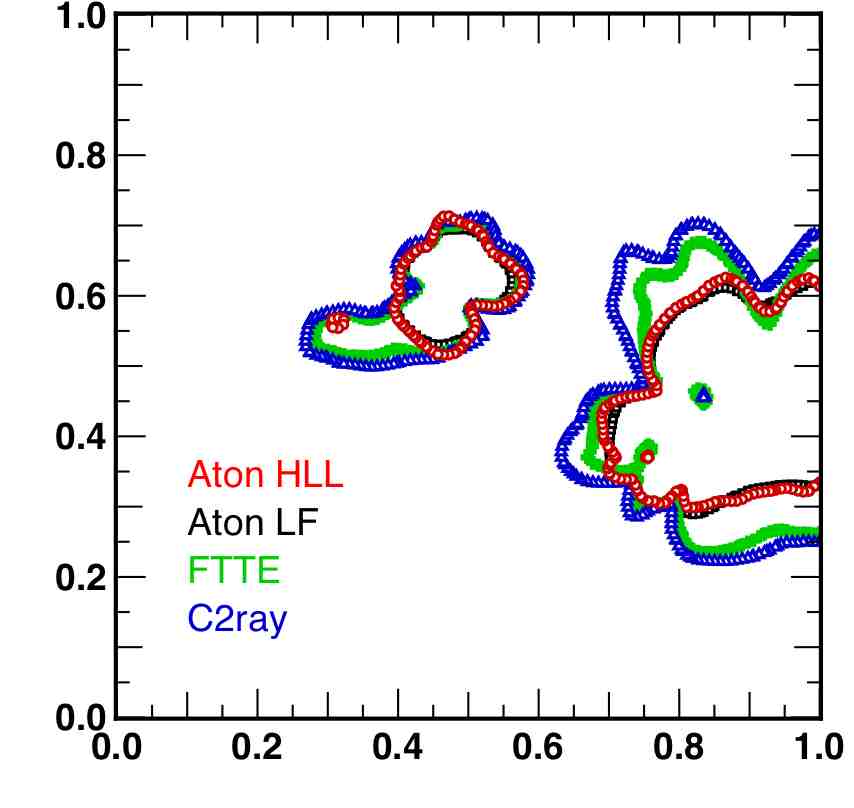}}
\resizebox{0.8\columnwidth}{0.8\columnwidth}{\includegraphics{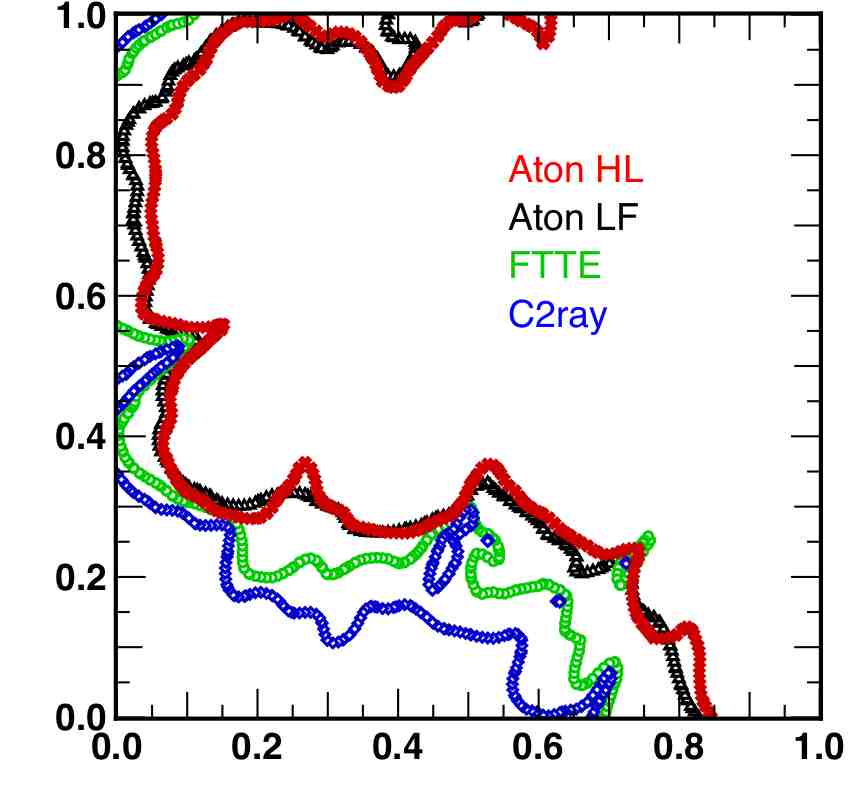}}
\caption{Static cosmological field test. I-front positions (defined as
a 50$\%$ ionized fraction) computed at 0.05 Myr (\textit{Top}) and 0.4
Myr  (\textit{bottom})  after  the  sources  were  ignited.  the  four
calculations were made using ATON-GLF, ATON-HLL, C2RAY and FTTE.}
\label{f:ifrontcosmo} 
\end{figure}

\begin{figure} 
\centering
\resizebox{0.8\columnwidth}{0.75\columnwidth}{\includegraphics{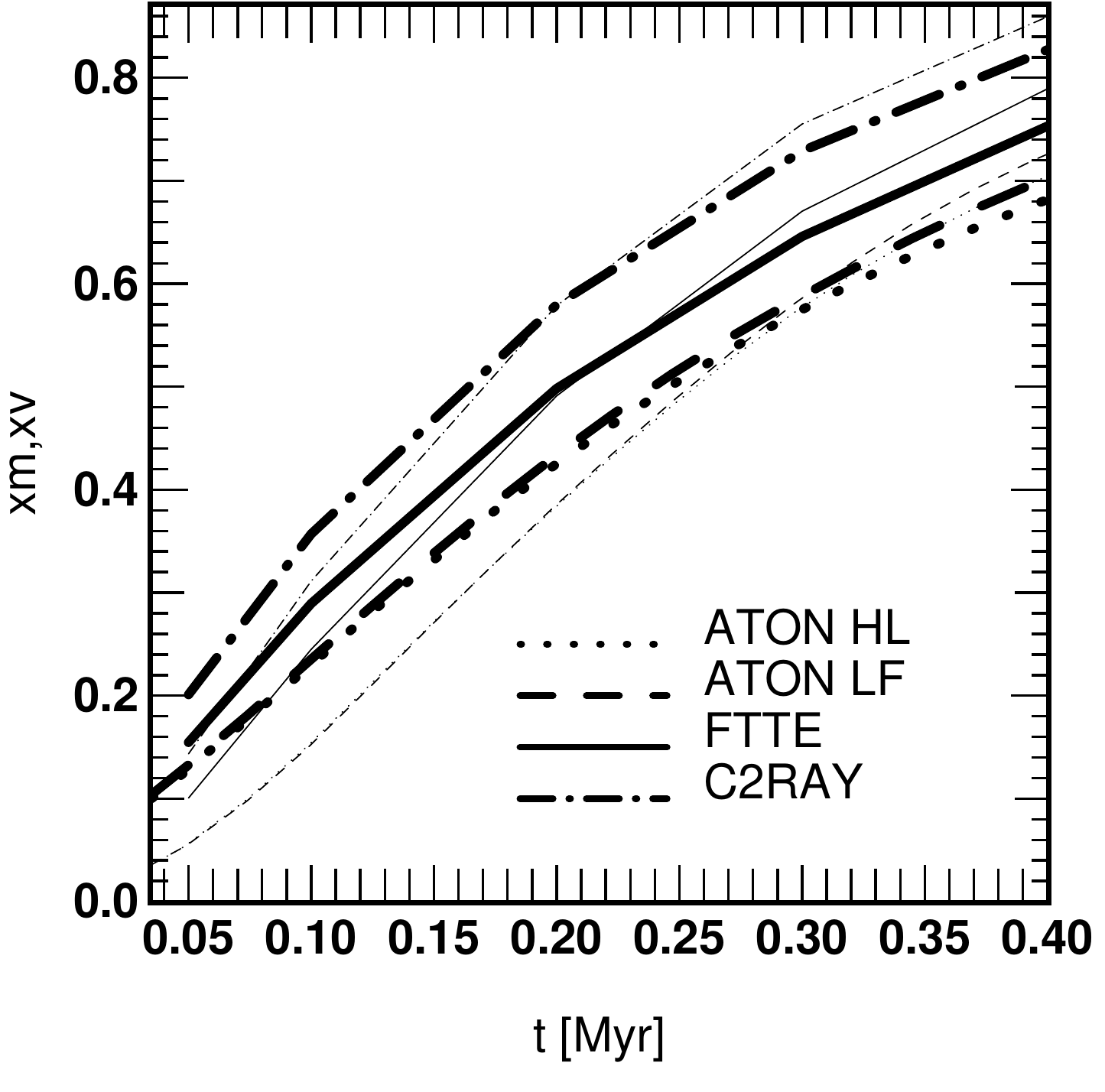}}
\resizebox{0.8\columnwidth}{0.75\columnwidth}{\includegraphics{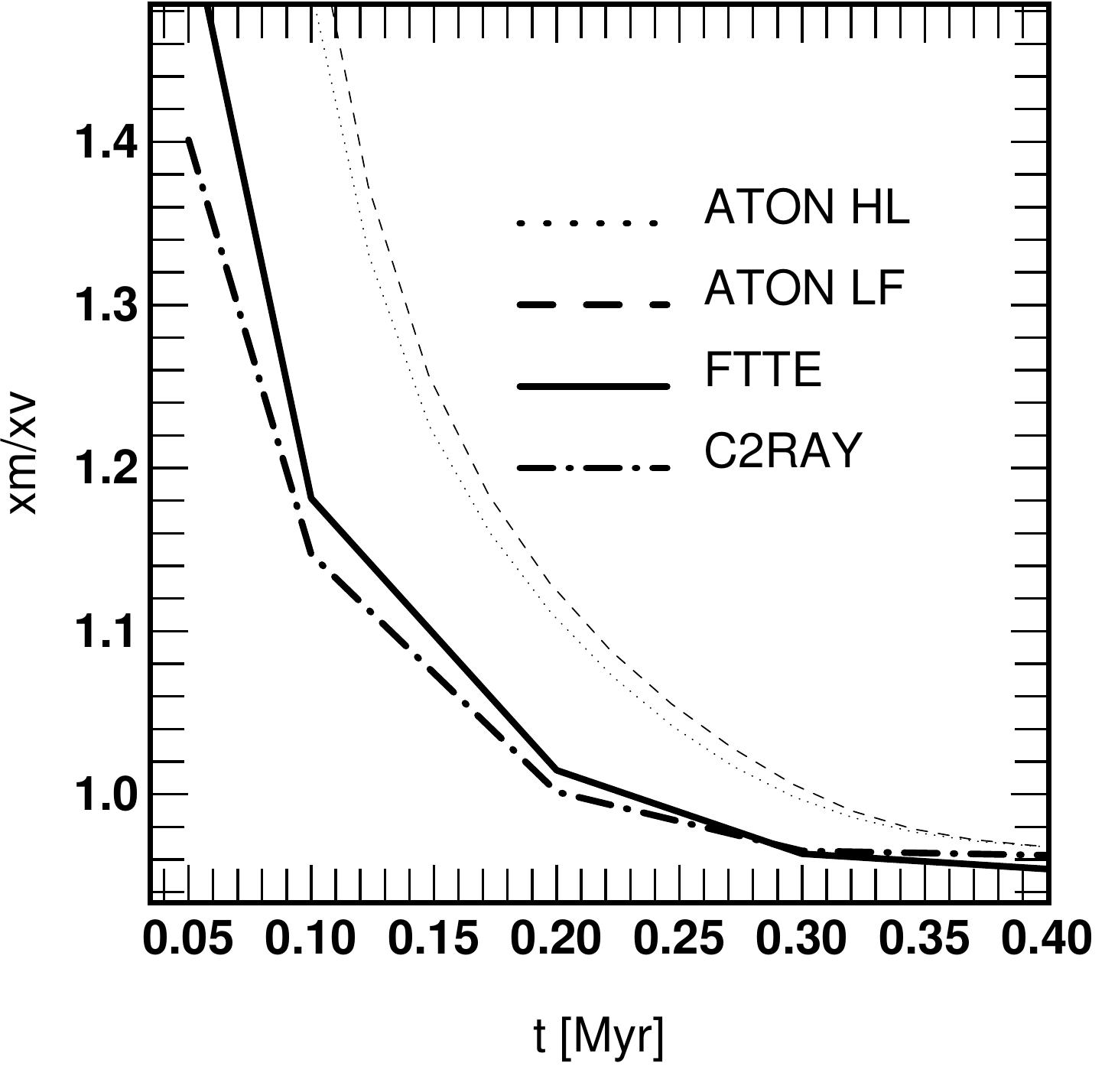}}
\caption{Static cosmological field  test. \textit{Top~:} the evolution
of the average ionized fraction $x_{v}$ (thin lines) and mass-weighted
average  $x_{m}$  (thick lines).  These  quantities  were computed  on
snapshots     produced     by     C2RAY,     FTTE,     ATON-GLF     and
ATON-HLL. \textit{Bottom~:}  the evolution of  the ratio $x_{m}/x_{v}$
for the same four calculations. }
\label{f:xmxvcosmo} 
\end{figure}

\begin{figure} 
\centering
\resizebox{0.8\columnwidth}{0.8\columnwidth}{\includegraphics{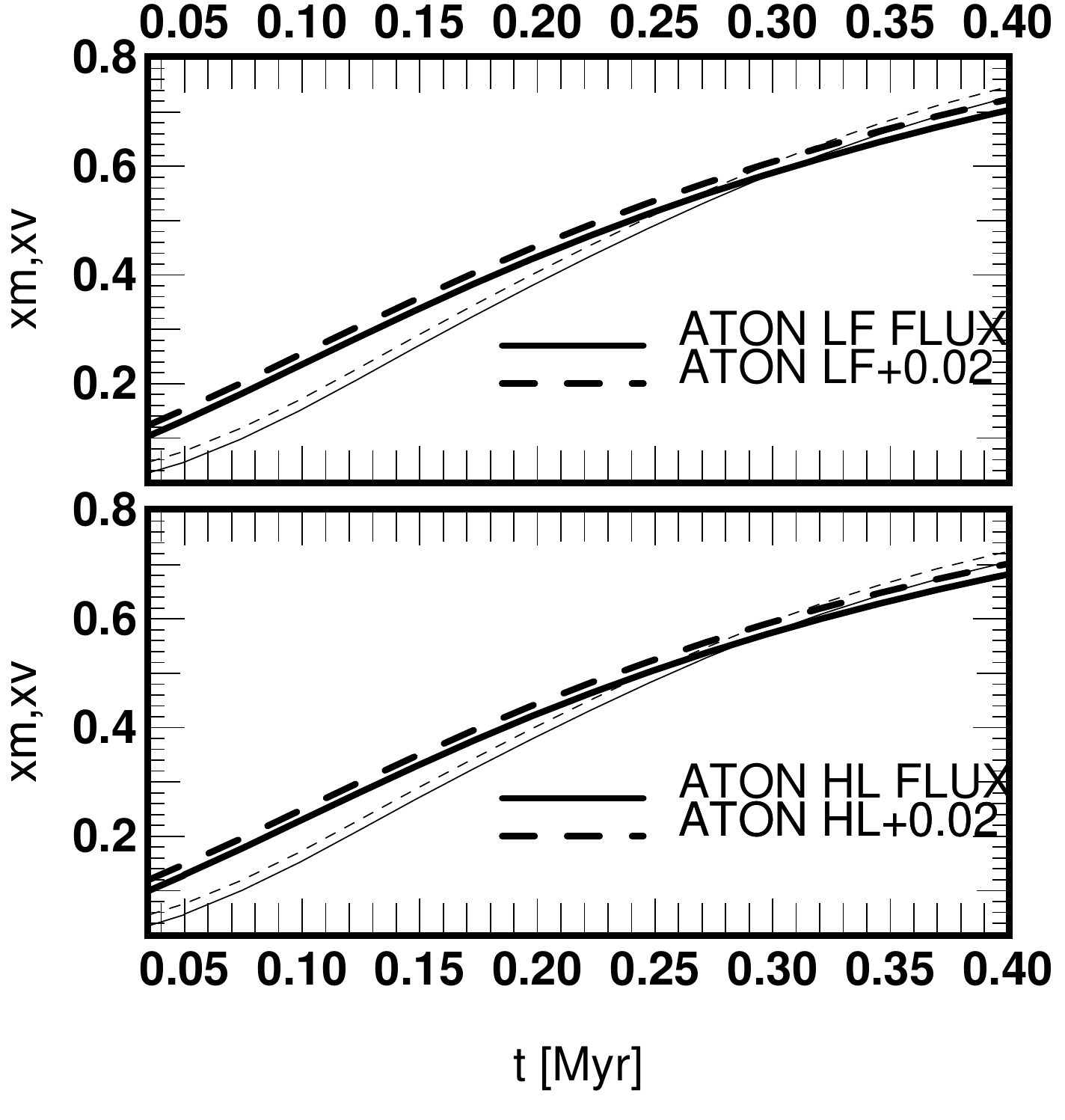}}
\caption{Static cosmological field test.  The evolution of the average
ionized  fraction  $x_{v}$  (thin  lines)  and  mass-weighted  average
$x_{m}$ (thick lines) using the regular scheme and the ``ray--tracing''
scheme  (with \textit{flux}  mention).  An offset  of  0.02 has  been
applied to  the latter quantity.  GLF (top) and HLL  (bottom) intercell
flux were  used. Clearly the  ``ray--tracing''  scheme do not  lead to
significant changes in the global evolution of the ionization.}
\label{f:xmxvcosmo2} 
\end{figure}

The final test consists in  modeling the propagation of I-fronts in an
heterogeneous  medium with  multiple  sources. The  gas density  field
consists in  a single snapshot  extracted from a  $128^3$ cosmological
simulation. The comoving box length is $500h^{-1}$ kpc and for sake of
simplicity  only the  $z=8.9$ snapshot  is  taken in  account and  the
medium is considered  as being static, both from the  point of view of
cosmology (e.g. no overall expansion) or local motion. A simple source
model and  distribution was also  derived, based on the  properties of
the biggest haloes. All the sources  are turned on a the same time and
have $10^5$K black-body spectrum. The initial temperature is 100 K and
the  simulation  is ran  over  400 000  years,  i.e.  well before  the
stationary  regime  achieved. ATON's  computations  are  made with  an
effective speed of light equal to $c$. Further details can be found in
\citet{2006MNRAS.371.1057I}.

\Fig{mapcosmox} and \Fig{mapcosmox2}  present the neutral fraction and
the temperature map at t=0.05 Myr and t=0.4 Myr~: only the mid-section
planes of the simulation are  shown. These maps were obtained with the
two  versions of  ATON (GLF  and HLL)  and compared  to the  same tests
performed    by   FTTE    (\citet{2005MNRAS.362.1413R})    and   C2RAY
(\citet{2006NewA...11..374M})        as       part        of       the
\citet{2006MNRAS.371.1057I} comparison project.  Let us emphasize that
the simulation  is run  over 400  000 years only,  i.e $0.1\%$  of the
recombination time and the regime  investigated in this test is highly
non-stationary.  Clear   differences  can  be   noted.  First,  ATON's
calculations seem to present a delay in the ionization propagation. It
can be seen from both maps where ionized and hot regions are typically
less extended  around the sources  in ATON's calculations  compared to
the other results. Second, hot  regions are more tightly correlated to
ionized regions in  ATON's scheme, while FTTE and  C2ray show extended
and complex  structures of hot  gas. Since the transition  between hot
and  cooler regions  is more  abrupt  in the  ATON's calculations,  it
relates to  the more compact  ionized regions that the  current scheme
predicts. Furthermore,  let us recall that  the current implementation
of  the scheme is  "monochromatic", even  though the  average photon's
energy reflects the spectrum of  the sources. As a consequence, fronts
are sharper  and high-energy photons  with large travel  distances are
not  available and  cannot  preheat  and preionize  the  gas on  large
scales.

Still,  given  this strong  restriction  to  a single  frequency-group
treatment,  results  can  be  seen as  qualitatively  satisfying.  The
I-front  positions  (see \Fig{ifrontcosmo})  with  ATON share  similar
features with the other codes.  For instance they coincides along high
density regions  such as filaments  axes. Densities are such  that the
frequency dependence  of cross-section do  not penalize the  scheme at
the current resolution compared to the others and conversely, I-fronts
in  voids  are  "late"  in  ATON:  this supports  the  fact  that  the
differences are due to the current simplified spectral treatment. This
is particularly evident in the  $0.05$ Myr map, where a good agreement
coincides  with   filaments  and  a  poor   agreement  coincides  with
voids. The $0.4$ Myr maps still hint such correlations but they appear
as less obvious since it results from a longer evolution.

The overall ionization evolution  is presented in \Fig{xmxvcosmo}. The
average ionized fraction is defined by
\begin{equation}
x_{v}=\frac{\Sigma_i^\mathrm{N} x_{i}}{\mathrm{N}},
\end{equation}
where $N$ stands  for the total number of grid cells,  $i$ is the cell
index and the mass-weighted average ionized fraction by
\begin{equation}
x_{m}=\frac{\Sigma_i^\mathrm{N} \rho_{i}x_{i}}{\Sigma_i^\mathrm{N} \rho_{i}},
\end{equation}
where $\rho_{i}$ is  the local gas density within  cell $i$. The ratio
of these two quantities is equal  to the ratio between the average gas
density    of    ionized    regions    to    the    average    density
(\citet{2006MNRAS.369.1625I})~:
\begin{equation}
\frac{x_{m}}{x_{v}}=\frac{M_{\mathrm{ionized}}}{{\bar \rho}V_{\mathrm{ionized}}}.
\end{equation}
A ratio  larger than one  implies that ionized regions  have densities
larger than the  average. Conversely a ratio smaller  than unity means
that voids dominate the  'population' of ionized regions.  Compared to
the  other  codes calculations,  ATON  exhibits  an  overall shift  of
$x_{m}(t)$  and  $x_{v}(t)$  in  amplitude ($\sim4\%$  with  FTTE  and
$\sim12\%$ with  C2RAY) while the  global trends are similar.  One can
notice that for all the codes the curves for $x_{m}$ and $x_{v}$ cross
each  other at  some point,  implying  that high  density regions  are
ionized  first  while voids  tend  to  be  ionized later.  The  switch
operates later in ATON and confirms the code's slowness in low density
regions.  Still  the discrepancy with other codes  remains limited and
is  encouraging  in the  prospect  of  a  more complete  treatment  of
multi-frequency transfer.  Let us also emphasize that  the current set
up investigate a  highly transient regime and the  previous tests have
shown that a much better agreement is expected on longer time scales.

Finally,   all   these   tests   were   performed   using   also   our
``ray--tracing'' scheme and no significant differences were found (see
\Fig{xmxvcosmo2}). The averaged  ionized fractions remain unchanged at
the percent level. The maps and the i-front positions (not shown here)
are quasi-identical. It implies that even though the isotropic recombination is artificially suppressed, it does
not  have a  strong  impact  compared to  the  lack of  multi frequency
treatment for large scale calculations. In the context of cosmological
reionization,  future  developments  would nevertheless  focus  on  these
aspects first.

\section{Discussion \& Conclusion}
ATON  is a Eulerian  scheme  for  radiative transfer  that  relies on  a
momentum description of the radiative transfer equation. The hierarchy
set up by  the conservation equations of energy  and radiative flux is
closed by means  of a relation between the  radiative pressure and the
energy. The 'M1' closure relation has been retained~: it expresses the
Eddington tensor as the  combination of a pure transport configuration
of the radiation and a  purely diffusive geometry. In the intermediate
cases ,  the relative  contribution of these  two regimes  is obtained
from the local flux properties.   This scheme is complemented with a a
simple  treatment  of the  local  chemistry,  to  compute the  neutral
hydrogen  abundance, and the  conservation of  energy, to  compute the
photoheating/cooling.  At  the   current  stage,  the  multi-frequency
treatment has not been implemented,  even though there are no a priori
difficulties for such an extension.  ATON is a simple scheme that will
eventually be  used to investigate  astrophysical ionization processes
such as the cosmological reionization.

ATON  has   been  tested   following  the  experiments   suggested  by
\citet{2006MNRAS.371.1057I}:  the propagation of  an HII  region, with
and without  auto-consistent heating, the  shadowing by a  dense clump
and the  I-front propagation in  a static cosmological field.  Most of
the results  obtained by ATON  are in agreement with  the calculations
presented in this  article. The main differences arises  from the lack
of multi-frequency  treatment~: because the spectrum  hardening is not
taken  in account, ionization  and heating  processes occurs  on small
scales  and for  instance no  large distance  pre-heating due  to high
energy photons is being modeled at the moment. It results on a loss of
radiative energy on small scales and I-fronts tend to be slower in low
density  regions.  However, it  only  causes  problems  in highly  non
stationary regimes while ATON catches all the details of the processes
in situations where I-front propagation are slowed down. The issue of the suppression of isotropic recombination was also assessed. For this purpose, we used a modified scheme that mimics ray-tracing codes by taking in account an anisotropic source of flux due to recombination~:   the  results obtained  are  very  similar  to ray-tracing  or Monte-Carlo  codes. However,  little differences were observed in  the
cosmological test, emphasizing the greater impact of the mono-group treatment.
 
%The intrinsic
%diffusivity of the  scheme was also assessed and  appears to be larger
%than other types of calculations.  A way to cope with this diffusivity
%is presented and is tested on the shadowing tests: using this enhanced
%scheme,  the  results obtained  are  very  similar  to ray-tracing  or
%Monte-Carlo  codes. 

Among  the  current  and  future  developments,  the  multi-frequency
treatment will  follow in  order to investigate  accurately ionization
processes  in highly  non  stationary regimes.  Such situations  would
occur on small scales close to the sources, and these studies would be
valuable to constrain e.g. the  escape fraction on scales smaller than
the resolution of  large volume simulations. Also, ATON  is limited to
the post-processing  of simulations and lacks the  coupling that exist
between  the dynamic  of the  gas and  the radiation.  Because  of its
Eulerian nature,  the current scheme  can be easily coupled  with grid
based hydrodynamical  codes and it  is planned to  be part of  the AMR
cosmological code RAMSES \citep{2002A&A...385..337T}.

Among  the  astrophysical  applications  of  ATON, the  study  of  the
cosmological  reinsertion is a  primary objective.  The post-processing 
of the AMR and SPH version of a large ongoing  hydrodynamical simulation is on the
way and will allow to compare the impact of the source distribution
and the  gas geometry  on  the reionization
process generated by each  version . It would  also lead to predictions on the  geometry of the 21
cm  emission,  in  the  prospects   of  experiments  such  as  SKA  or
LOFAR. Comparisons on  the propagation of radiation in  and around the
biggest objects,  at high resolution are  also on the way  in order to
investigate the impact  of small scales clustering of  the gas and the
star   distribution.  Hopefully,   with  the   improvements  mentioned
previously, the  current scheme  will be able  to investigate  all the
astrophysical processes  where radiation is  relevant, while remaining
simple to conceive and implement.\\

{\bf Acknowledgments}

{\sl The authors would like to thank E. Audit, I. Iliev and B. Semelin
for valuable comments and discussions. A special thanks to M. Gonzalez
for providing  us his  routine to compute  the eigenvalues for  the M1
Model.  We also like to thank the Cosmological Radiative Transfer Comparison Project for making their data publicly available (http://www.cita.utoronto.ca/~iliev/rtwiki/doku.php). This  work was performed  with support from the  {\em HORIZON}
project (http://www.projet-horizon.fr). }

\bibliography{aton}

\begin{thebibliography}{}

\bibitem[\protect\citeauthoryear{{Barkana} \& {Loeb}}{{Barkana} \&
  {Loeb}}{2001}]{2001PhR...349..125B}
{Barkana} R.,  {Loeb} A.,  2001, \physrep, 349, 125

\bibitem[\protect\citeauthoryear{{Dubroca} \& {Feugeas}}{{Dubroca} \&
  {Feugeas}}{1999}]{M1}
{Dubroca} B.,  {Feugeas} J.,  1999, C. R. Acad. Sci., 329, 915

\bibitem[\protect\citeauthoryear{{Gnedin} \& {Abel}}{{Gnedin} \&
  {Abel}}{2001}]{2001NewA....6..437G}
{Gnedin} N.~Y.,  {Abel} T.,  2001, New Astronomy, 6, 437

\bibitem[\protect\citeauthoryear{{Gonz{\'a}lez}, {Audit} \&
  {Huynh}}{{Gonz{\'a}lez} et~al.}{2007}]{2007A&A...464..429G}
{Gonz{\'a}lez} M.,  {Audit} E.,    {Huynh} P.,  2007, \aap, 464, 429

\bibitem[\protect\citeauthoryear{{Hui} \& {Gnedin}}{{Hui} \&
  {Gnedin}}{1997}]{1997MNRAS.292...27H}
{Hui} L.,  {Gnedin} N.~Y.,  1997, \mnras, 292, 27

\bibitem[\protect\citeauthoryear{{Iliev}, {Ciardi}, {Alvarez}, {Maselli},
  {Ferrara}, {Gnedin}, {Mellema}, {Nakamoto}, {Norman}, {Razoumov},
  {Rijkhorst}, {Ritzerveld}, {Shapiro}, {Susa}, {Umemura} \& {Whalen}}{{Iliev}
  et~al.}{2006}]{2006MNRAS.371.1057I}
{Iliev} I.~T.,  {Ciardi} B.,  {Alvarez} M.~A.,  {Maselli} A.,  {Ferrara} A.,
  {Gnedin} N.~Y.,  {Mellema} G.,  {Nakamoto} T.,  {Norman} M.~L.,  {Razoumov}
  A.~O.,  {Rijkhorst} E.-J.,  {Ritzerveld} J.,  {Shapiro} P.~R.,  {Susa} H.,
  {Umemura} M.,    {Whalen} D.~J.,  2006, \mnras, 371, 1057

\bibitem[\protect\citeauthoryear{{Iliev}, {Mellema}, {Pen}, {Merz}, {Shapiro}
  \& {Alvarez}}{{Iliev} et~al.}{2006}]{2006MNRAS.369.1625I}
{Iliev} I.~T.,  {Mellema} G.,  {Pen} U.-L.,  {Merz} H.,  {Shapiro} P.~R.,
  {Alvarez} M.~A.,  2006, \mnras, 369, 1625

\bibitem[\protect\citeauthoryear{{Katz}, {Weinberg} \& {Hernquist}}{{Katz}
  et~al.}{1996}]{1996ApJS..105...19K}
{Katz} N.,  {Weinberg} D.~H.,    {Hernquist} L.,  1996, \apjs, 105, 19

\bibitem[\protect\citeauthoryear{{Levermore}}{{Levermore}}{1984}]{1984JQSRT..3%
1..149L}
{Levermore} C.~D.,  1984, Journal of Quantitative Spectroscopy and Radiative
  Transfer, 31, 149

\bibitem[\protect\citeauthoryear{{Madau}, {Haardt} \& {Rees}}{{Madau}
  et~al.}{1999}]{1999ApJ...514..648M}
{Madau} P.,  {Haardt} F.,    {Rees} M.~J.,  1999, \apj, 514, 648

\bibitem[\protect\citeauthoryear{{Maselli}, {Ferrara} \& {Ciardi}}{{Maselli}
  et~al.}{2003}]{2003MNRAS.345..379M}
{Maselli} A.,  {Ferrara} A.,    {Ciardi} B.,  2003, \mnras, 345, 379

\bibitem[\protect\citeauthoryear{{Mellema}, {Iliev}, {Alvarez} \&
  {Shapiro}}{{Mellema} et~al.}{2006}]{2006NewA...11..374M}
{Mellema} G.,  {Iliev} I.~T.,  {Alvarez} M.~A.,    {Shapiro} P.~R.,  2006, New
  Astronomy, 11, 374

\bibitem[\protect\citeauthoryear{{Mellema}, {Iliev}, {Pen} \&
  {Shapiro}}{{Mellema} et~al.}{2006}]{2006MNRAS.372..679M}
{Mellema} G.,  {Iliev} I.~T.,  {Pen} U.-L.,    {Shapiro} P.~R.,  2006, \mnras,
  372, 679

\bibitem[\protect\citeauthoryear{{Razoumov} \& {Cardall}}{{Razoumov} \&
  {Cardall}}{2005}]{2005MNRAS.362.1413R}
{Razoumov} A.~O.,  {Cardall} C.~Y.,  2005, \mnras, 362, 1413

\bibitem[\protect\citeauthoryear{{Teyssier}}{{Teyssier}}{2002}]{2002A&A...385.%
.337T}
{Teyssier} R.,  2002, \aap, 385, 337

\bibitem[\protect\citeauthoryear{{Toro}}{{Toro}}{1999}]{Toro97}
{Toro} E.,  1999, {Riemann Solvers and Numerical Methods for Fluid Dynamics}.
Berlin, Germany, Springer-Verlag, 1999, 624 p.

\end{thebibliography}
\bibliographystyle{mn2e}

\appendix

\section{Implicit computation of ionization fraction}
In ATON, the coupling between radiative energy and ionization fraction
is  solved implicitly.  
%In this  appendix, the  details of  the method
%implemented  are described.  
Let us  call  $N_{\gamma}$ the  density number  of
photons (i.e. the  energy density in single photon  units), $F_{\gamma}$ the
associated flux, $P_{\gamma}$ the radiative pressure and  $x$ the ionization fraction. Let  us also define
$\rho$ as the initial  hydrogen density and $\alpha$, $\alpha_{B}$ and
$\beta$ as respectively the case  A and case B recombination rates and
the collisional ionization rate. First we consider a \textit{fixed} temperature (from the previous timestep) for the rates computation. 
The  set of coupled (1D) equations is given by
\begin{eqnarray}
\frac{\md N_{\gamma}}{\md t}+\frac{\md F_{\gamma}}{\md r}&=&\dot N_{\gamma}^*+\dot N_{\gamma}^{\mathrm{rec}}-\rho\sigma_{\gamma} cN_{\gamma}(1-x)\label{e:e2}\\
\frac{\md F_{\gamma}}{\md t}+\frac{\md P_{\gamma}}{\md r}&=&-\rho\sigma_{\gamma} c F_{\gamma}.
\end{eqnarray}
Here $\dot N_{\gamma}^*$  and $\dot N_{\gamma}^{\mathrm{rec}}$ stands for  the point-like (namely  the 'stars') and
the diffuse  source of  \textit{ionizing} photons due to recombination and  $\sigma_{\gamma}$ stands
for   the  photo-ionization  cross-section   averaged  over   the  $N_{\gamma}$
spectrum. The ionization equation is given by
\begin{equation}
\frac{\md \rho (1-x)}{\md t}=\alpha\rho^2 x^2-\beta x(1-x)\rho^2-\rho(1-x)N_{\gamma}\sigma_{\gamma} c.\label{e:e3}
\end{equation}
The equation over $n$ can be rewritten as:
\begin{eqnarray}
\frac{\md N_{\gamma}}{\md t}+\frac{\md F_{\gamma}}{\md r}&=&S -\alpha_{B}\rho^2 x^2 +\beta x(1-x) \rho^2-\rho\frac{\md x}{\md t},\label{e:e1}
\end{eqnarray}
where  we  assumed that  $\alpha\rho^2  x^2=\dot N_{\gamma}^{\mathrm{rec}}+\alpha_{B}\rho ^2  x^2$.
With  $p$  labeling  the  time  step index,  we  write  ${\md  N_{\gamma}}/{\md
t}=(N_{\gamma}^{p+1}-N_{\gamma}^{p})/\Delta  t$  and  $x=x^p$  and $X=x^{p+1}$  and  the
implicit formulation of \Eq{e1} is given by:
\begin{eqnarray}
N_{\gamma}^{p+1}=N_{\gamma}{'}+\beta\rho^2(1-X)X\Delta t -\alpha_{B}\rho^2X^2\Delta t -\rho(X-x),\label{e:e4}
\end{eqnarray}
where $N_{\gamma}'$  is the  explicit solution of  the pure  advection equation
(i.e. \Eq{e2} with a zero r.h.s.) given by:
\begin{equation}
N_{\gamma}'=\dot N_{\gamma}^*\Delta t -\frac{\md F_{\gamma}}{\md r}\Delta t+N_{\gamma}^p.
\end{equation}
Combining \Eq{e2} and \Eq{e4}, one can obtain a third-order polynomial
$Q_{3}(X)=mX^3+nX^2+pX+q$, where the coefficients are given by
\begin{eqnarray}
m&=&(\alpha_{B}+\beta)\rho^2\Delta t\\
n&=&\rho-(\alpha+\beta)\rho/\sigma_{\gamma} c-\alpha_{B}\rho^2\Delta t-2\beta\rho^2\Delta t\\
p&=&-\rho(1+x)-N_{\gamma}'-1/\sigma_{\gamma} c \Delta t+\beta\rho/\sigma_{\gamma} c+\beta\rho^2\Delta t\\
q&=&N_{\gamma}'+\rho x +x/\sigma_{\gamma} c\Delta t,
\end{eqnarray}
and where  $X$ is  given by finding  the roots of  $Q_{3}(X)$. Knowing
$X$, $N_{\gamma}$ is obtained from \Eq{e4} and $F_{\gamma}$ can be derived from
\begin{equation}
F_{\gamma}^{p+1}=\frac{F_{\gamma}'}{1+c\sigma_{\gamma}\rho\Delta t(1-X)},
\end{equation}
where  $F_{\gamma} '$ is  the explicit  solution of  the the  pure advection
equation of the flux, given by:
\begin{equation}
F_{\gamma}'=F_{\gamma}^p-\frac{\md P_{\gamma}}{\md r}\Delta t.
\end{equation}

The   temperature  at   time  $t+\Delta   t$  is   obtained  \textit{a
posteriori}, having computed values of $N_{\gamma}^{p+1}$, $F_{\gamma}^{p+1}$ and
$x^{p+1}$  while  using  the  temperature  set at  the  previous  time
step. From \Eq{temp}, the (discrete) equation that rules the temperature's evolution is given by:  
\begin{equation}
\frac{T^{p+1}-T^p}{\Delta t}=\frac{2({\cal H-L}-\frac{3}{2}\rho(1+X) k_{B} T^{p+1} (X-x)/\Delta t)}{3\rho(1+X)k_{B}}\label{e:dt},
\end{equation}
$\cal H$ and $\cal L$ being the heating and cooling rates.
Therefore, \Eq{dt}  can  be solved  separately  in an  explicit
manner where the R.H.S depends on $T^p$ instead of $T^{p+1}$.
%\begin{equation}
%\frac{T^{p+1}-T^p}{\Delta t}=\frac{3(H-\Lambda-\frac{3}{2}\rho_0(1+x^{p+1}) k T^p (x^{p+1}-x^p)/\Delta t)}{2\rho_0(1+x^{p+1})k}.
%\label{e:dtemp}
%\end{equation}
This equation is  purely local  and is  solved at  each cell's
location   without   any  spatial   coupling.   By  simple   algebraic
manipulation, the  updated value of  the temperature $T^{p+1}$  can in
principle be  easily obtained. However, the cooling  time becomes much
shorter than  $\Delta t$ as  temperature reach typical value  of $\sim
10^4$  K, therefore \Eq{dt}  cannot be  solved without  resolving this
typical time scale.

In practice, \Eq{dt} is sub-cycled during a radiative time step, using
$\Delta t =0.9 t_{\mathrm cool}$. At each 'temperature' iteration, the
cooling  rate  is updated,  setting  a new  $\Delta  t$  for the  next
iteration.   As  a  consequence  such a  procedure  can  substantially
increase the computing time. This  increase can be reduced by stopping
the sub-cycling  when some relative convergence of  the temperature is
achieved.  In the  tests presented  hereafter, a  condition such  as a
$10^{-6}$ convergence after at least $100$ sub-cycles is found to give
satisfying result in terms of ionization fraction distributions. Finally the whole process is repeated: the ionized fraction is re-computed using the updated temperature until a global convergence is achieved.

%ATON's   updating   scheme   relies   on   the   following
%discretization of the energy and ionization equations:
%\begin{eqnarray}
%\frac{n^{p+1}-n^{p}}{\Delta t}&=&-\rho_0\frac{X-x}{\Delta t}-\nabla{\bf \Phi}-\alpha_{b}\rho_0^2X^2+\Sigma^p\label{e:degy}\\
%\frac{X-x}{\Delta t}&=&-\alpha X^2\rho_0+n^{p+1}\langle\sigma\rangle c(1-X)\label{e:dxion}\\
%%\frac{T^{p+1}-T^p}{\Delta t}&=&\frac{2(H-\Lambda-\frac{3}{2}\rho_0(1+X) k T^{p+1} (X-x)/\Delta t)}{3\rho_0(1+X)k}\label{e:dt}.
%\end{eqnarray}
%Here $x$ and $X$ stands for the ionized fraction at time steps $p$ and
%$p+1$. One should note that  recombination and cooling rates depend on
%temperature. This set  of equation is purely local, i.e.  it has to be
%solved  at  each  cell's  location  and does  not  involve  a  spatial
%coupling.  The term $\alpha_{b}\rho_0^2X^2$  arises from  \Eq{ion}, by
%rewriting (omitting the collisional ionization):
%\begin{equation}
%n\langle\sigma\rangle c(1-x)=\alpha\rho_0^2x^2+\rho_0\frac{\md x}{\md t},
%\end{equation}
%where the  number of  recombinations satisfies the  following equality
%$\alpha \rho_0  x^2=\delta+\alpha_{b}\rho_0 x^2$, i.e is  equal to the
%sum of the  photons emitted by a diffuse  source of ionizing radiation
%plus non-ionizing (case B) photons. The  latter acts as a sink term in
%\Eq{dxion} (see Appendix for more details).

This model is admittedly oversimplified and a full implicit treatment
of  ionization and  photoheating would  be preferable.  However, it appears from  subsequent  experiments that  taking in account
the complexity  of the energy-ionization coupling with  a temperature
that varies in background  returns satisfying  results. Since
temperature  is essentially  crucial to  compute  recombination rates,
which do not depend strongly on temperature in our case, this simple model appears
to be accurate enough at the current stage.

\end{document}